\documentclass[12pt]{article}
\usepackage{amssymb,amsmath,amsthm,latexsym,natbib,amsbsy,mathtools}
\usepackage{graphicx}
\usepackage{float}
\usepackage{caption}
\usepackage{subcaption}
\usepackage{epsfig}
\usepackage{epstopdf}  % For using .eps files when compiling directly to .pdf
\usepackage{color}
\usepackage[font= {small,it} ]{caption}
\usepackage{url}
\usepackage{doi}
\usepackage[titletoc]{appendix}
\usepackage{xcolor}  % To change the color of text
\usepackage[boxed]{algorithm2e} % for algorithms

\newcommand{\blind}{0}  % Controls blinding
\newcommand{\B}{\boldsymbol}

\graphicspath{"}

\date{}

\begin{document}
\def\spacingset#1{\renewcommand{\baselinestretch}%
{#1}\small\normalsize} \spacingset{1}

%%%%%%%%%%%%%%%%%%%%%%%%%%%%%%%%%%%%%%%%%%%%%%%%%%%%%%%%%%%%%%%%%%%%%%%%%%%%%%

\if0\blind
{
  \title{\bf Sampling Strategies for Fast Updating of Gaussian Markov Random Fields}
  \author{D. Andrew Brown\thanks{
    D. Andrew Brown is Assistant Professor in the School of Mathematical and Statistical Sciences at Clemson University, Clemson, SC 29634 (email: ab7@clemson.edu). Christopher S. McMahan is Associate Professor in the School of Mathematical and Statistical Sciences at Clemson University. Stella Watson Self is a PhD Candidate in the School of Mathematical and Statistical Sciences at Clemson University. This material is based upon work partially supported by the National Science Foundation (NSF) under Grant DMS-1127974 to the Statistical and Applied Mathematical Sciences Institute. DAB is partially supported by NSF Grants CMMI-1563435, EEC-1744497 and OIA-1826715. CSM is partially supported by National Institutes of Health Grant R01 AI121351 and NSF grant OIA-1826715.}\hspace{.2cm}\\
    School of Mathematical and Statistical Sciences, Clemson University\\
    Clemson, SC, USA 29634-0975\\
    and \\
    Christopher S. McMahan\\
    School of Mathematical and Statistical Sciences, Clemson University\\
    and \\
    Stella Watson Self \\
    School of Mathematical and Statistical Sciences, Clemson University}
  \maketitle
} \fi

\if1\blind
{
  \bigskip
  \bigskip
  \bigskip
  \begin{center}
    {\LARGE\bf Sampling Strategies for Fast Updating of Gaussian Markov Random Fields}
\end{center}
  \medskip
} \fi
%\maketitle

%\thispagestyle{empty}

\begin{abstract}
\noindent Gaussian Markov random fields (GMRFs) are popular for modeling dependence in large areal datasets due to their ease of interpretation and computational convenience afforded by the sparse precision matrices needed for random variable generation. Typically in Bayesian computation, GMRFs are updated jointly in a block Gibbs sampler or componentwise in a single-site sampler via the full conditional distributions. The former approach can speed convergence by updating correlated variables all at once, while the latter avoids solving large matrices. We consider a sampling approach in which the underlying graph can be cut so that conditionally independent sites are updated simultaneously. This algorithm allows a practitioner to parallelize updates of subsets of locations or to take advantage of `vectorized' calculations in a high-level language such as \verb|R|. Through both simulated and real data, we demonstrate computational savings that can be achieved versus both single-site and block updating, regardless of whether the data are on a regular or an irregular lattice. The approach provides a good compromise between statistical and computational efficiency and is accessible to statisticians without expertise in numerical analysis or advanced computing.\\
%\begin{keywords}
%Bayesian computation, Cholesky factorization, chromatic Gibbs sampling, conditional autoregressive model, graph coloring, Markov chain Monte Carlo
%\end{keywords}
\end{abstract}

\noindent%
{\it Keywords:}  Bayesian computation, Cholesky factorization, chromatic Gibbs sampling, conditional autoregressive model, graph coloring, Markov chain Monte Carlo
\vfill

\newpage
\spacingset{1.45} % DON'T change the spacing!

\section{INTRODUCTION}
Suppose we have observed data $\B{y} = (y_1, \ldots, y_n)^T$ in which each $y_i$ summarizes information over an area $i, ~i= 1, \ldots, n$, such as a sum or average of individuals in the area. For instance, \citet{SelfEtAl18} investigate regional trends of occurrence of Lyme disease, where the data are the number of positive disease cases observed in each county in the United States. Other examples include \cite{BrownEtAl14}, who consider functional magnetic resonance imaging data in which each $y_i$ quantifies the neuronal changes associated with an experiment observed in the $i^{\text{th}}$ three-dimensional pixel in a brain image, where the goal is to identify those areas exhibiting statistically significant changes. \citet{WallerEtAl97} estimate spatially-varying risks of developing lung cancer using reported deaths in each county of the state of Ohio. In the examples we consider in this work, $y_i$ is either the observed intensity at pixel $i$ in an image or the number of votes cast for a particular candidate in voting precinct $i$ in the state of New York. The task in the former is to reconstruct an underlying true image that has been corrupted with noise; in the latter we aim to estimate spatially-varying trends in voter preference throughout the state.

What these examples, and countless others, have in common is that the data are correlated so that the value at one location is influenced by the values at nearby locations. While this dependence can be directly modeled in the likelihood of $\B{y}$, it is often reasonable to assume that it can be explained by an unobservable process $\B{x} = (x_1, \ldots, x_n)^T$, where $x_i$ is the realization of the process at node (location) $i$. Then a typical Bayesian analysis of this problem takes the $y_i$'s to be {\em conditionally} independent given $\B{x}$; $y_i \mid \B{x} \stackrel{\text{indep.}}{\sim} f_i(\cdot \mid \B{x}), ~i= 1, \ldots, n$. In other words, the correlation is assumed to be completely explained by $\B{x}$. For more flexibility and to more fully account for sources of uncertainty, one might assume that the distribution of $\B{x}$ is determined by an unknown parameter vector $\B{\theta}$ (usually of much smaller dimension than $\B{x}$) which is itself assigned a hyper-prior. Thus, the Bayesian model is \begin{equation}\label{eq:genBayesMod}
    \begin{aligned}
        \B{y} \mid \B{x} &\sim f(\cdot \mid \B{x})\\
        \B{x} \mid \B{\theta} &\sim \pi_x(\cdot \mid \B{\theta})\\
        \B{\theta} &\sim \pi_\theta(\cdot).\\
    \end{aligned}
\end{equation}
Inference proceeds by evaluating (or estimating) characteristics of the posterior distribution, determined via Bayes' rule as $\pi(\B{x}, \B{\theta} \mid \B{y}) \propto f(\B{y} \mid \B{x})\pi_x(\B{x} \mid \B{\theta})\pi_\theta(\B{\theta})$.

A widely adopted approach for modeling the dependence structure in this problem is to assume $\B{x}$ satisfies a Markov property. In the simplest case, this means that if $x_j$ is in between $x_i$ and $x_k$, then $x_i$ and $x_k$ are conditionally independent, given $x_j$. (Higher-order neighborhoods are also sometimes used where conditioning on more values is necessary.) If $\B{x}$ satisfies this property, then $\B{x}$ is said to be a {\em Markov random field} (MRF). MRFs are useful tools in a variety of challenging applications, including disease mapping \citep{WallerEtAl97, SelfEtAl18}, medical imaging \citep{Higdon98, BrownEtAl14}, and gene microarray analysis \citep{XiaoEtAl09, BrownEtAl17}. Even autoregressive time series models are instances of Markov random fields; though this work is primarily motivated by models for spatially-indexed data in which there is no clear direction of influence. Awareness of such models was raised after the seminal work of \citet{Besag74}, after which they came to be known in the statistics literature as {\em conditional autoregressive} \citep[CAR;][]{BanerjeeEtAl15} models. Since then, they have become popular for modeling temporally- or spatially-dependent areal data due to their interpretability and computational tractability afforded by the conditional independence induced by the Markov property. This property is particularly important for modern Markov chain Monte Carlo \citep[MCMC;][]{GelfandSmith90} methods. Indeed, the ease with which Markov random fields can be incorporated into a Gibbs sampling algorithm \citep{GemanGeman84} has contributed to their popularity in Bayesian statistics.

We are concerned in this work with models in which $\B{x} \mid \B{\theta}$ is a {\em Gaussian} Markov random field. Gaussian Markov random fields \citep[GMRFs;][]{RueHeld05} are simply MRFs in which the conditional distribution of each (scalar) random variable is Gaussian. GMRFs typically are specified either implicitly by providing the complete set of full conditional distributions $p(x_i \mid x_1, \ldots, x_{i-1}, x_{i+1}, \ldots, x_n), ~i= 1, \ldots, n$, or explicitly by defining the precision (inverse covariance) matrix instead of the covariance function as would be done in Gaussian process modeling \citep{SchabenGotway05}. Further, GMRFs do not usually yield stationary processes due to a so-called ``edge effect" in which the marginal variances vary by location. Corrections can be made to yield a stationary process such as a periodic boundary assumption \citep{FoxNorton16} or algorithmic specification of the precision matrix \citep{Dempster72}. Sometimes the effect can simply be ignored with little effect on inference \citep{BesagKoop95}. Efforts have been made to use GMRFs to approximate Gaussian processes with specified covariance functions \cite[e.g.,][]{RueTjem02, SongEtAl08, LindgrenEtAl11}, but much work still remains.

A particularly intuitive instance of a GMRF is one that centers the distribution of each $x_i$ at the average of its neighbors; i.e., $x_i \mid \B{x}_{(-i)} \sim N(\bar{x}_i, \sigma_i)$, where $\B{x}_{(-i)} = (x_1, \ldots, x_{i-1}, x_{i+1}, \ldots, x_n)^T$, $\bar{x}_i$ is the average of the values adjacent to $x_i$, and $\sigma_i^2$ is obtained by scaling a common variance term by the number of neighbors at site $i$. The precision matrix determined by this model is only positive {\em semi-}definite and thus not invertible, meaning that the joint distribution is improper. Such models are called {\em intrinsic autoregressive} \citep[IAR;][]{BesagKoop95} models and are popular as Bayesian prior distributions, due in part to their interpretability.

% Brief overview of sampling problem (literature review?)
Belonging to the Gaussian class of distributions, GMRFs are the most widely studied Markov random fields. See \cite{RueHeld05} for an overview of relevant work. The literature includes techniques for efficiently sampling from GMRFs. As we discuss in Section 2, the two most common methods for sampling both have caveats when working with extremely high-dimensional data. So-called block sampling involves Cholesky factorizations of large precision matrices and thus carries high computational and memory costs. While a GMRF prior induces sparsity which can be exploited to economize such calculations, conditional posterior precision matrices arising in Bayesian models such as $\eqref{eq:genBayesMod}$ typically depend on parameters that change in each iteration of an MCMC algorithm and the required repeated factorizations can be extremely time consuming. On the other hand, so-called single-site samplers work by only considering scalar random variable updates. In addition to being more loop-intensive than block samplers, single-site samplers are known to exhibit slow convergence when the variables are highly correlated \citep{CarlinLouis09}. The competing goals of statistical efficiency and computational efficiency have led to recent innovations in alternative sampling approaches for GMRFs. Some of these approaches require considerable expertise in numerical analysis or message passing interface (MPI) protocol, but others are relatively easy to implement and hence can be quite useful for statisticians. Specifically, the recently proposed chromatic Gibbs sampler \citep{GonzalezEtAl11} is easy to implement and is competitive with or even able to improve upon other existing strategies. It allows a practitioner to parallelize sampling or to take advantage of `vectorized' calculations in a high-level language such as \texttt{R} \citep{R16}  without requiring extensive expertise in numerical analysis or MPI.

The chromatic sampler appearing in \cite{GonzalezEtAl11} was motivated by and demonstrated on binary MRFs. However, it is straightforward to carry over the same idea to the Gaussian case. In this paper, we discuss block updating and single-site updating of GMRFs and compare them to chromatic sampling. Rather than focusing on theoretical convergence rates or an otherwise overall ``best" approach, we view these techniques through the lens of a practitioner looking for easily implemented yet efficient algorithms.  To the best of our knowledge, this work is the first time chromatic Gibbs sampling has been directly compared to the standard approaches for sampling of GMRFs.

There exist fast approximation methods for estimating features of a posterior distribution without resorting to Markov chain Monte Carlo. One of the most popular of these is integrated nested Laplace approximation \citep[INLA;][]{RueEtAl09}, the \verb|R| implementation of which is the \verb|R-INLA| package \citep{LindgrenRue15}. Such approximation methods are useful when certain quantities need to be estimated quickly, but they are only approximations and thus are not interchangeable with Markov chain Monte Carlo algorithms that converge to the exact target distribution and allow for the approximation of virtually any posterior expected value with the same Monte Carlo sample. Indeed, INLA provides the most accurate approximations around the posterior median and can disagree with MCMC in tail probability approximations \citep{GerberFurrer15}. These disagreements are more pronounced in cases where the full conditional distribution of the random field is non-Gaussian (for which INLA uses Laplace approximations) and a GMRF is used as a proposal in a Metropolis-Hastings algorithm. Further, the \verb|R-INLA| package is a ``black box" that works well for a set of pre-defined models. For more flexibility to manipulate non-standard models, there is the need to break open the black box to customize an algorithm to suit one's needs. In the context of GMRFs, this requires more direct interaction with the random fields, motivating this work. Efficient strategies such as those considered here are not intended to be substitutes for INLA or other approximation methods. Rather, they are complementary procedures that are useful when one is interested in direct MCMC on challenging posterior distributions.

In Section \ref{sec:HGS}, we briefly motivate our sampling problem and review GMRFs. We then compare chromatic sampling to block updating and single-site sampling of GMRFs. In Section \ref{sec:sims} we compare the performance of single-site sampling, block updating, and the chromatic approach in a numerical study using a simple Bayesian model with spatial random effects on simulated, high-dimensional imaging data, as well as a real application involving non-Gaussian polling data. We conclude in Section \ref{sec:disc} with a discussion.

\section{MCMC SAMPLING FOR GAUSSIAN MARKOV RANDOM FIELDS}\label{sec:HGS}

In modern Bayesian analysis, it is common for the posterior distribution to have no known closed form. Hence, expectations with respect to this distribution cannot be evaluated directly. If one can obtain a sample from this distribution, though, laws of large numbers allow us to approximate quantities of interest via Monte Carlo methods. A common approach to obtaining a sample from a posterior distribution is Markov chain Monte Carlo (MCMC), particularly Gibbs sampling.

One reason for the popularity of Gibbs sampling is the ease with which the algorithm can be constructed. For an estimand $\B{\mu} = (\mu_1, \ldots, \mu_p)^T$, it proceeds simply by initializing a chain at $(\mu_1^{(0)}, \ldots, \mu_p^{(0)})^T$ and, at iteration $t$, sampling $\mu_m^{(t)} \sim \pi(\mu_m \mid \mu_1^{(t)}, \ldots, \mu_{m-1}^{(t)}, \mu_{m+1}^{(t-1)}, \ldots, \mu_p^{(t-1)}),$ $m= 1, \ldots, p$. Under suitable conditions, ergodic theory \citep[e.g.,][]{RobertCasella04} establishes that the resulting Markov chain $\{(\mu_1^{(t)}, \ldots, \mu_p^{(t)})^T: t= 0, 1, \cdots\}$ has $\pi(\B{\mu})$ as its limiting distribution. In practice, for GMRFs with target distribution $\pi(\B{x}, \B{\theta} \mid \B{y})$, implementing this algorithm requires the ability to draw $\B{x} \mid \B{\theta}, \B{y}$ thousands of times. This is computationally expensive and thus quite challenging when $\B{x}$ is high dimensional, as we discuss in this Section.

\subsection{Gaussian Markov Random Fields}
Consider a GMRF $\B{x} = (x_1, \ldots, x_n)^T$, where $x_i$ is the realization of the field at node $i, ~i= 1, \ldots n$. The density of $\B{x}$ is given by
\begin{eqnarray}\label{eqn:GMRFDens}
    \begin{aligned}
        \pi(\B{x} \mid \B{\mu})
            &\propto \exp\left(-\frac{1}{2}\B{x}^T\B{Q}\B{x} + \B{b}^T\B{x}\right),\\
    \end{aligned}
\end{eqnarray}
where $\B{\mu} \in \mathbb{R}^n$ and $\B{b} = \B{Q}\B{\mu}$. If $\B{Q}$ is nonsingular, then this distribution is proper (i.e., $\int \pi(\B{x} \mid \B{\mu}) d\B{x} < \infty,$ for all $\B{\mu}$) and the normalizing constant is $(2\pi)^{-n/2}\text{det}(\B{Q})^{1/2}$. Intrinsic GMRFs are such that $\B{Q}$ is rank deficient and only positive semidefinite. In this case, we may define the density with proportionality constant $(2\pi)^{-(n-k)/2}\text{det}^*(\B{Q})^{1/2}$, where $n-k$ is the rank of $\B{Q}$ and $\text{det}^*(\cdot)$ is the product of the $n-k$ non-zero eigenvalues of $\B{Q}$ \citep{HodgesEtAl03, RueHeld05}. Such improper GMRF models are common in Bayesian disease mapping \citep{WallerEtAl97} and linear inverse problems \citep{Bardsley12}, as they are easily interpretable and usually yield proper posterior distributions.

An appealing feature of GMRFs is the ability to specify the distribution of $\B{x}$ through a complete set of full conditional distributions, $\{p(x_i \mid \B{x}_{(-i)}) : ~i= 1, \ldots, n\}$. For instance, we can assume each $x_i \mid \B{x}_{(-i)} \sim N(\eta_i, \sigma^2_i)$, with $\eta_i = \mu_i + \sum_{j \sim i}c_{ij}(x_j - \mu_j)$ and $\sigma_i^2 > 0$, where $i \sim j$ if and only if node $i$ is connected to (i.e., a neighbor of) node $j \neq i$ and $c_{ij}$ are specified weights such that $c_{ij} \neq 0$ if and only if $i \sim j$. Specification of a Markov random field through these so-called {\em local characteristics} was pioneered by \citet{Besag74}, after which such models came to be known as conditional autoregressive (CAR) models. \cite{Besag74} uses Brook's Lemma and the Hammersley-Clifford Theorem to establish that the set of full conditionals collectively determine a joint density, provided a positivity condition holds among $\B{x}$. In this case, we have that $\B{Q}_{ij} = (I(i = j) - c_{ij}I(i \neq j))/\sigma_i^2$, where $I(\cdot)$ is the indicator function. The condition $\sigma_j^2c_{ij} = \sigma_i^2c_{ji}$, for all $i,j$, is necessary to ensure symmetry of $\B{Q}$. The ease with which these full conditional distributions can be incorporated into a Gibbs sampling algorithm has led to a dramatic increase in the popularity of CAR models over the past twenty years or so \citep{Lee2013, BanerjeeEtAl15}. Indeed, there exists user-friendly software that facilitates incorporating CAR models into Bayesian spatial models without detailed knowledge of their construction. Examples include the \verb|GeoBUGS| package in \verb|WinBUGS| \citep{LunnEtAl00} and the \verb|R| package \verb|CARBayes| \citep{Lee2013}.

GMRFs may be specified according to an undirected graph $\mathcal{G} = (\mathcal{V}, \mathcal{E}),$ where $\mathcal{V}$ indicates nodes (the vertices) and $\mathcal{E} = \{(i,j) : i \sim j\}$ is the edge set. The precision matrix $\B{Q}$ is determined by $(\B{Q})_{ij} \neq 0$ if and only if $(i,j) \in \mathcal{E}$. Specifying the density through the precision matrix $\B{Q}$ instead of a covariance matrix induces a Markov property in the random field \citep[][Theorem 2.2]{RueHeld05}. For any node $i$, $x_i \mid \B{x}_{(-i)} \stackrel{d}{=} x_i \mid \B{x}_{\mathcal{N}(i)}$, where $\mathcal{N}(i) = \{j : (i,j) \in \mathcal{E}$\} is the neighborhood of node $i$ and $\B{x}_{\B{A}} := (x_i\ : i \in \B{A})^T$ for some index set $\B{A}$. That is, $x_i$ is conditionally independent of the rest of the field given its neighbors. Most GMRFs assume that each node has relatively few neighbors, resulting in $\B{Q}$ being sparse. The typical sparsity of the precision matrix  is another reason GMRFs are widely used to model dependence in areal data.

With the need to model extremely large datasets with nontrivial correlation has come the need for efficient sampling techniques whereby posterior distributions arising from fully Bayesian models can be simulated. When periodic boundary conditions on $\B{x} \in \mathbb{R}^n$ can be assumed (i.e., each $x_i$ has the same number of neighbors, including the edge nodes, as in a pixelized image with zero-padded boundaries), \citet{FoxNorton16} note that the sampling problem can be diagonalized via the Fast Fourier Transform (with complexity $\mathcal{O}(n\log n)$), whence a sample can be drawn by solving a system in $\mathcal{O}(n)$ operations. They propose reducing the total number of draws from the conditional distribution of $\B{x}$ by using a ``marginal-then-conditional" sampler in which the MCMC algorithm operates by completely collapsing over $\B{x}$ and subsequently sampling $\B{x}$ using only the approximately independent draws of the hyperparameters obtained from a full MCMC run on their marginal distribution. In many applications, though, the periodic boundary assumption may not be realistic (e.g., administrative data indexed by irregular geographic regions or pathways in microarray analysis consisting of different numbers of genes), and sampling from the marginal distribution of hyperparameters can itself be challenging. To avoid the computational difficulties associated with full GMRFs, \citet{CaiEtAl13} propose using a pairwise graphical model as an approximate GMRF for high-dimensional data imputation without specifying the precision matrix directly. The authors admit, however, that this procedure is very hard to implement \citep[][p. 7]{Cai14}. In cases where we are given $\B{Q}$ and $\B{b}$ in (\ref{eqn:GMRFDens}) with the goal of estimating $\B{\mu}$, \citet{JohnsonEtAl13} express the Gibbs sampler as a Gauss-Seidel iterative solution to $\B{Q\mu} = \B{b}$, facilitating the ``Hogwild" parallel algorithm of \citet{NiuEtAl11} in which multiple nodes are updated simultaneously without locking the remaining nodes. In the Gaussian case, \citet{JohnsonEtAl13} prove convergence to the correct solution when the precision matrix $\B{Q}$ is symmetric diagonally dominant. Motivated by \citet{JohnsonEtAl13}, \citet{ChengEtAl15} use results from spectral graph theory to propose a parallel algorithm for approximating a set of sparse factors of $\B{Q}^l, ~-1 \leq l \leq 1,$ in nearly linear time. They show that it can be used to construct independent and identically distributed realizations from an approximate distribution. This is opposed to a Gibbs sampler, which produces approximately independent samples from the correct distribution (possibly after thinning). Similar to the Gauss-Seidel splitting considered by \citet{JohnsonEtAl13}, \citet{LiuEtAl15} propose an iterative approach to approximating a draw from a GMRF in which the corresponding graph is separated into a spanning tree and the missing edges, whence the spanning tree is randomly perturbed and used as the basis for an iterative linear solve.

The aforementioned algorithms can be difficult to implement and require substantial knowledge of graph theory, numerical analysis, and MPI programming. This makes such approaches inaccessible to many statisticians who nevertheless need to work with large random fields. Further, they are iterative routines for producing a single draw from an approximation to the target distribution. This feature makes them less appealing for users who work in \verb|R| or \verb|MATLAB| (The MathWorks, Natick, MA). It is well-known that loops should be avoided in these languages to avoid repeated data type interpretation and memory overhead issues. In response to these difficulties while still faced with the problem of efficient updating of GMRFs inside a larger MCMC algorithm, additional \verb|R| packages have been made available which are beneficial for manipulating the sparse matrices associated with GMRFs, including \verb|Matrix| \citep{BatesMaechlerMatPackage}, \verb|SparseM| \citep{KoenkerNgSparseM}, and \verb|spam| \citep{FurrerSainSpam, GerberFurrer15}.

\subsection{Block and Single-Site Gibbs Sampling}\label{ssec:singleSite}
In this Section, it is helpful to distinguish between sampling $\B{x}$ directly from a {\em prior} GMRF and from the full conditional distribution of $\B{x}$ derived from an hierarchical Bayesian model with a GMRF prior on $\B{x}$. For an unconditional (and proper) GMRF, the distribution is of the form $\B{x} \sim N(\B{\mu}, \B{Q}^{-1})$, where $\B{\mu}$ and $\B{Q}$ are generally unrelated. When drawing from the full conditional distribution as in a Gibbs sampler, the distribution is of the form $\B{x} \sim N(\B{Q}_p^{-1}\B{b}, \B{Q}_p^{-1}),$
%\begin{equation}\label{eqn:condGauss}
%    \B{x} \sim N(\B{Q}_p^{-1}\B{b}, \B{Q}_p^{-1}),
%\end{equation}
where $\B{Q}_p \neq \B{Q}$ is an updated precision matrix. For example, in a typical linear model $\B{y} \mid \B{x}, \B{\Sigma} \sim N(\B{Ax}, \B{\Sigma})$ with $\B{A}$ fixed and $\B{x} \sim N(\B{\mu}, \B{Q}^{-1})$, standard multivariate normal theory yields $\B{x} \mid \B{y}, \B{\Sigma} \sim N(\B{Q}_p^{-1}\B{b}, \B{Q}_p^{-1})$, where $\B{Q}_p = \B{A}^T\B{\Sigma}^{-1}\B{A} + \B{Q}$ and $\B{b} = \B{A}^T\B{\Sigma}^{-1}\B{y} + \B{Q\mu}$.

Two approaches to updating GMRFs inside a Markov chain Monte Carlo algorithm are so-called {\em single site sampling} in which individual sites are updated one at a time using the available full conditional distributions, and {\em block Gibbs sampling} in which the entire random field is updated all at once via sampling from a known multivariate Gaussian distribution induced by the GMRF. Block sampling improves the convergence of Gibbs samplers in the presence of {\em a posteriori} correlated variables by allowing the chain to move more quickly through its support \citep{LiuEtAl94}. The drawback is in the manipulation and solution of large covariance matrices necessary for both random variable generation and evaluation of the likelihood in a Metropolis-Hastings algorithm \citep{MetropolisEtAl53, Hastings70}. Single site updating uses the conditional distributions of each scalar random variable, thus avoiding large matrix computations. In single-site sampling, though, statistical efficiency may be sacrificed as updating a group of possibly highly correlated parameters one at a time can result in slow exploration of the support, slowing convergence of the Markov chain.

In single-site Gibbs sampling, we sequentially draw from each univariate conditional distribution with density $p(x_i \mid \B{x}_{(-i)}), ~i= 1,\ldots n$. Broadly speaking, this requires alternating $n$ times between using $\B{x}_{(-i)} \in \mathbb{R}^{n-1}$ to calculate the conditional mean and drawing from $p(x_i \mid \B{x}_{(-i)})$, meaning that single-site updating essentially is an $\mathcal{O}(n^2)$ operation. blueHowever, by exploiting the fact that the conditional mean (usually) only depends on a relatively few neighbors of $x_i$ (i.e., $\mathcal{N}(i) \ll n$), updating it after each draw becomes negligible, reducing the complexity to $\mathcal{O}(n)$. This algorithm has little regard for the ordering of the nodes, making such sampling strategies very easy to implement. Compared to block updating, though, many more Gibbs scans may be required to sufficiently explore the support of the distribution. This approach is the most iteration-intensive of any of the approaches considered here. As such, its implementation in \verb|R| can result in a large amount of overhead associated with loops, considerably slowing the entire routine.

Efficient block sampling schemes for GMRFs are discussed in \cite{Rue01} and \cite{KnorrHeldRue02}. What most of these schemes have in common is the use of a Cholesky factorization to solve a system of equations.  For the case typically encountered in a Gibbs sampler, \cite{RueHeld05} provide an algorithm for simulating from $N(\B{Q}_p^{-1}\B{b}, \B{Q}_p^{-1})$. This algorithm, presented in Algorithm 1 in the Supplementary Material, requires one Cholesky factorization and three linear solves via forward or backward substitution.

In general, for a matrix of dimension $n \times n$, the Cholesky factorization is an $\mathcal{O}(n^3/3)$ operation and each linear solve costs $\mathcal{O}(n^2)$ flops \citep{GolubVanLoan96}. This can be particularly onerous in a fully Bayesian approach in which hyperpriors are assigned to hyperparameters $\B{\theta}$ that appear in the precision matrix $\B{Q}_p \equiv \B{Q}_p(\B{\theta})$. However, the key to making block updating feasible on high-dimensional data lies in the computational savings that can be achieved when $\B{Q}_p$ is sparse. Sparse matrix algebra is itself a non-trivial problem requiring specialized knowledge beyond the expertise of many statisticians. Indeed, concerning this point, \citet[][p. 52]{RueHeld05} recommend ``leaving the issue of constructing and implementing algorithms for factorizing sparse matrices to the numerical and computer science experts." In practice, most statisticians rely on special functions for sparse matrices such as those found in the \verb|Matrix|, \verb|SparseM|, or \verb|spam|  packages in \verb|R|. Of these three, \verb|spam| is the most specifically tailored for repeatedly sampling GMRFs in MCMC.

For simulating posterior distributions via block Gibbs sampling, we are interested in drawing from full conditional distributions. In this case, sparsity of the entire precision matrix is contingent upon the sparsity of $\B{A}^T\B{\Sigma}^{-1}\B{A}$. This is often the case in practice. For instance, in disease mapping and related applications, it is common to place a spatially correlated random effect at each location to encourage smoothing of the incidence rate over space \cite[e.g.,][]{WallerEtAl97, BanerjeeEtAl15}. In terms of the linear model, this can be expressed as $\B{y} - \B{X\beta} = \B{Z\gamma} + \B{\varepsilon}$, where $\B{X\beta}$ corresponds to fixed effects and $\B{\gamma}$ contains the spatially-varying effects. With site-specific random effects, $\B{Z}$ is diagonal or block diagonal. The diagonal case (e.g., $\B{Z} = \B{I}$) is especially amenable to efficient block Gibbs sampling as well as chromatic sampling (see Subsection \ref{ssec:HGS}), since the underlying graph $\mathcal{G}$ for the full conditional distribution is exactly the same as the prior graph.

For a fully Bayesian model, the full conditional precision matrix associated with the GMRF will generally depend on parameters that are updated in each iteration of an MCMC routine, meaning that the Cholesky factorization has to be recomputed on each iteration. Often, though, the neighborhood structure and thus the sparsity pattern of the Cholesky factor remain fixed. The sparse matrix implementation in the \verb|spam| package exploits this fact to accelerate repeated block GMRF updates. After finding the initial Cholesky factorization using so-called supernodal elimination trees \citep{NgPetyon93}, \verb|spam| stores the symbolic factorization and only performs numeric factorizations on subsequent iterations. Even with sparse matrix algebra, though, block sampling $\B{x}$ from a GMRF in very high dimensions can be problematic due to the computational cost of even an initial factorization as well as the associated memory overhead \citep[][p. 331]{Rue01}.

\subsection{Chromatic Gibbs Sampling}\label{ssec:HGS}
Consider the graph representation of the GMRF, $\mathcal{G} = (\mathcal{V}, \mathcal{E})$. The {\em local Markov property} says that $x_i \perp \B{x}_{-(i, \mathcal{N}(i))} \mid \B{x}_{\mathcal{N}(i)}$, where $\B{x}_{-(i, \mathcal{N}(i))}$ denotes all $x$ except $x_i$ and the neighborhood of $x_i$, and $\perp$ denotes (statistical) independence. An extension of the local Markov property is to let $C \subset \mathcal{V}$ denote a separating set, or {\em cut}, of $\mathcal{G}$ such that nodes in a set $A \subset \mathcal{V}$ are disconnected from nodes in $B \subset \mathcal{V}$ after removing the nodes in $C$ from the graph. Then the {\em global Markov property} states that $\B{x}_{A} \perp \B{x}_B \mid \B{x}_C$. Chromatic sampling exploits this property by partitioning the nodes according to a graph coloring whereby the nodes in each subset can be updated simultaneously.

\begin{figure}[tb]
    \centering
    % sizing = L B R T
    %\includegraphics[clip= TRUE, trim= 0.5in 7.5in 0.5in 0.75in, scale= 0.55]{"Revision/Revision Images/Figure".pdf}
    \includegraphics[clip= TRUE, trim= 0in 2.75in 0in 0.75in, scale= 0.35]{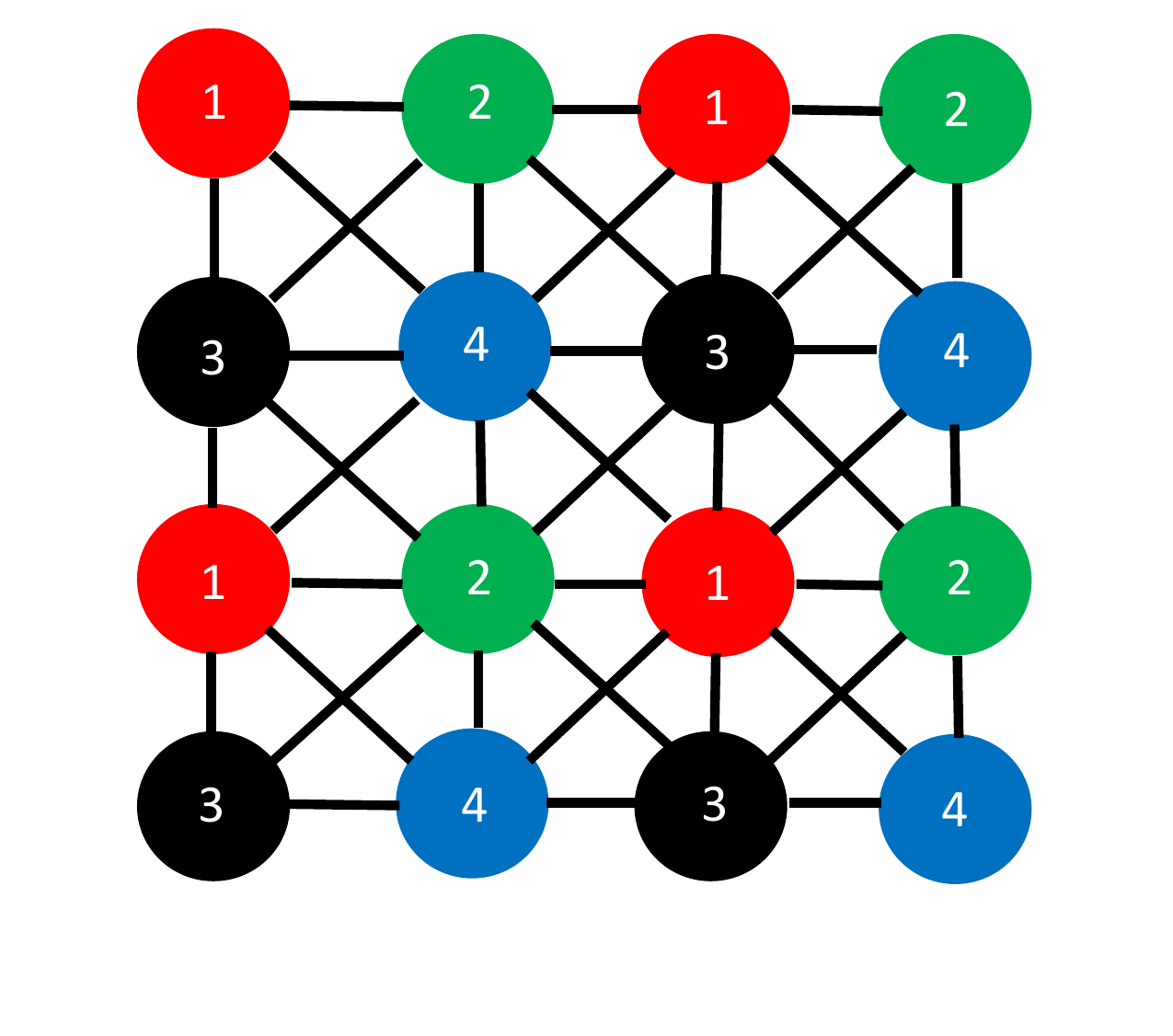}
    \caption{An example of a $k$-coloring ($k = 4$) for nodes on a regular two-dimensional lattice.}\label{fig:colEx}
\end{figure}

A {\em coloring} $f: \mathcal{V} \rightarrow \{1, \ldots, k\}, ~k \in \mathbb{N},$ is a collection of labels assigned to nodes on a graph so that no two nodes that share an edge have the same label. A $k$-coloring induces a partition of the nodes $\{\mathcal{A}_1, \ldots, \mathcal{A}_k\},$ where $\mathcal{A}_j = f^{-1}(\{j\}) \subset \mathcal{V}$. For example, Figure \ref{fig:colEx} displays a 4-coloring that could be used for data that lie on a regular two-dimensional lattice; e.g., imaging data. Given a $k$-coloring of the MRF graph, we can determine a cut $C_j$ corresponding to each color $j$ by assigning all nodes that are not of that color to be in the cut; i.e., $C_j = \mathcal{A}_j^c, ~j= 1, \ldots, k$. Defining cuts in this way for $j= 1, \ldots, k$, we have that $x_i \mid \B{x}_{C_j} \stackrel{\text{indep.}}{\sim} N(\eta_i, \sigma_i^2)$, for all $i \in \mathcal{A}_j,$
%\[
%    x_i \mid \B{x}_{C_j} \stackrel{\text{indep.}}{\sim} N(\eta_i, \sigma_i^2), ~~i \in %\mathcal{A}_j,
%\]
where each $\eta_i$ and $\sigma_i^2$ depend on $\B{x}_{C_j}$. That is, all nodes of the same color are conditionally independent and can be sampled in parallel, given the rest of the field. The use of graph colorings in this way lead \cite{GonzalezEtAl11} to term the approach {\em chromatic Gibbs sampling}.

Algorithm \ref{ALG:HGS} presents a general chromatic Gibbs sampler for GMRFs. An advantage of using this approach is in step 3 of the algorithm. When updates of the random variables indexed by $\mathcal{A}_j$ are distributed across several processors, the computational effort of updating the entire field can potentially be dramatically reduced, even compared to the approximate linear complexity obtained from sparse matrix factorization. Given $c$ processors and a $k$-coloring of a Markov random field over $n$ nodes, and assuming calculating conditional means is $\mathcal{O}(1)$, the chromatic Gibbs sampler generates a new sample in approximately $\mathcal{O}(n/c + k)$ operations \citep[][p. 326]{GonzalezEtAl11}. Observe that single-site Gibbs sampling can be obtained as a case of chromatic Gibbs sampling with $k = n$ colors.

\LinesNumbered
\begin{algorithm}[tb]
\SetAlgoLined
\DontPrintSemicolon
\SetKwInput{Input}{Input}
\SetKwInput{Output}{Output}
	\Input{Current state of GMRF, $\B{x}^{(t)}$, a $k$-coloring of the MRF graph, $\{\mathcal{A}_j : ~j= 1, \ldots, k\}$.}
	\Output{New draw $\B{x}^{(t+1)}$ from the GMRF.}
	\BlankLine
	\For{$j = 1$ to $k$} {
  	     For $i \in \mathcal{A}_j$, calculate conditional means and standard deviations $\eta_i, \sigma_i^2$ using $\B{x}_{\mathcal{A}_{1}}^{(t+1)}, \ldots, \B{x}_{\mathcal{A}_{j-1}}^{(t+1)}, \B{x}_{\mathcal{A}_{j+1}}^{(t)}, \ldots, \B{x}_{\mathcal{A}_{k}}^{(t)}$\;
         Draw $\B{x}_{\mathcal{A}_j} \sim N\left[(\eta_1, \ldots, \eta_{|\mathcal{A}_j|})^T, ~\text{diag}(\sigma_i^2, i= 1, \ldots, |\mathcal{A}_j|)\right]$ \;
    }
    {\bf Return} $\B{x}^{(t+1)}$ \;
\caption{Chromatic Gibbs step updating of a GMRF.}
\label{ALG:HGS}
\end{algorithm}

The best computational savings under chromatic sampling will be achieved by using the {\em chromatic index} for the coloring, defined as the minimum $k$ so that a $k$-coloring of $\mathcal{G}$ exists. The minimal coloring problem for a graph is NP-hard and thus very challenging except in simple situations. On regular lattices with commonly assumed neighborhood structures (e.g., Figure \ref{fig:colEx}), such colorings can be found by inspection without complicated algorithms. Coloring more general graphs is more involved. However, it is important to observe that for fixed sparsity patterns (and hence fixed Markov graphs) such as those we consider here, graph coloring is a pre-computation. It is only required to run the algorithm once prior to running MCMC.

A straightforward approach to graph coloring is the greedy algorithm, but it is known to generally produce suboptimal colorings. In fact, for random graphs in which any two vertices have probability 1/2 of sharing an edge, the greedy algorithm is known to asymptotically produce, on average, twice as many colors as necessary \citep{Grimmett75}. We illustrate this with an example in the Supplementary Material in which the greedy algorithm produces the optimal coloring under one permutation of the vertices, and over twice as many colors with another permutation. The sensitivity of the greedy algorithm to the ordering of the vertices was recognized by \citet{Culberson92}, who proposes an iterative approach in which the greedy algorithm is repeatedly applied to permutations of the vertices so that the optimal coloring can be better approximated. Beyond greedy algorithms, \citet{KragerEtAl98} cast the $k$-coloring problem as a semidefinite program and propose a randomized polynomial time algorithm for its solution. A full exposition of coloring algorithms is well beyond the scope of this work. However, our experience is that even the suboptimal colorings produced by the simple greedy algorithm are still able to facilitate vast computational improvements over block or single-site updating. As such, we provide in the Supplementary Material Algorithm 2, an easily-implemented greedy algorithm that is accessible to most statisticians looking for a quick way to color their MRF graph.  We also remark that there exist \texttt{R} packages containing functions for coloring graphs; e.g., the \texttt{MapColoring} package \citep{HunMapColor17} which implements the DSATUR algorithm \citep{Brelaz79}.

% \LinesNumbered
% \begin{algorithm}[tb]
% \SetAlgoLined
% \DontPrintSemicolon
% \SetKwInput{Input}{Input}
% \SetKwInput{Output}{Output}
% 	\Input{MRF graph $\mathcal{G} = (\mathcal{V}, \mathcal{E})$.}
% 	\Output{$k$-coloring partition $\{\mathcal{A}_1, \mathcal{A}_2, \ldots\, \mathcal{A}_k\}$, for some $k$.}
% 	\BlankLine
% 	Set $j=1$ and $\mathcal{A}_0=\emptyset$\;
% 	\While{$\mathcal{V} \backslash \cup_{l=0}^{j-1}\mathcal{A}_l \neq \emptyset$} {
% 	    $\mathcal{I}_j \leftarrow \mathcal{V} \backslash \cup_{l=0}^{j-1}\mathcal{A}_l$ \;
% 	    $\mathcal{A}_j \leftarrow \emptyset$ \;
% 	    \While{$|\mathcal{I}_j| > 0$}{
% 	        $i \leftarrow \min \mathcal{I}_j$\;
% 	        $\mathcal{A}_j \leftarrow \mathcal{A}_j \cup \{i\}$\;
% 	        $\mathcal{I}_j \leftarrow \mathcal{I}_j \backslash\left(\{i\} \cup \mathcal{N}(i)\right)$\;
% 	    }
% 	    $j \leftarrow j+1$\;
% 	}
% 	$k \leftarrow j-1$\;
%     {\bf Return} $\{\mathcal{A}_1,...,\mathcal{A}_k\}$ \;
% \caption{Greedy algorithm for $k-$coloring the nodes of an MRF graph.}
% \label{ALG:greedy}
% \end{algorithm}
% %Hence the chromatic Gibbs sampling approach should be accessible to practitioners who lack any deep knowledge of graph theory.

% For large-scale computing environments in which the number of available processors $p =\mathcal{O}(n)$, chromatic Gibbs updating takes $\mathcal{O}(k)$ operations, a dramatic improvement when $k \ll n$. For many practitioners, of course, such resources will be unavailable.
Most computers today have parallel processing capabilities, and any distributed processing over $c$ processors can reduce the computational burden by an approximate factor of $1/c$. Regardless of the number of processors available to the user, though, savings can still be realized when working in a high-level language such as \verb|R| by `vectorizing' the updating of the conditionally independent sets. Vectorizing still ultimately uses a \verb|for| loop on each set of nodes, but the loops are performed in a faster language such as  \verb|C| or \verb|Fortran|. It also minimizes the overhead associated with interpreting data types; i.e., vectorizing allows \verb|R| to interpret the data type only once for the entire vector instead of repeatedly for each element of the vector.

\section{NUMERICAL ILLUSTRATIONS}\label{sec:sims}
In this Section we compare chromatic sampling to block Gibbs and single-site sampling with both simulated data on large regular arrays and real, non-Gaussian (binomial) data on an irregular lattice. Our emphasis here is on ease of implementation for statisticians who may not be as comfortable with low-level programming as they are in \texttt{R}. Thus, most of our comparisons are done by examining the computational effort associated with programming entirely in \texttt{R}. We emphasize that computational improvements may be realized {\em without} direct parallel processing. We simply vectorize the simultaneous updating steps, thereby avoiding direct \verb|for| loops in \verb|R|. It is important to note that our implementation of chromatic sampling involves updating the means after each simultaneous draw via matrix-vector multiplication. The necessary matrices are stored in sparse format. Without sparse representations, the computational effort would be dramatically increased. Near the end of Subsection \ref{ssec:ImgRest}, we consider also a parallel implementation of the chromatic sampler in \texttt{R}, as well as what happens when the single-site updating step is done in \texttt{C++} rather than \texttt{R}. To implement the block Gibbs sampler, we use the \verb|spam| package \citep{FurrerSainSpam}, since it is specifically tailored for GMRFs inside MCMC routines by storing the sparsity structure for repeated use. To make the \verb|spam| functions as efficient as possible, we follow the authors' suggestion and turn off the symmetry check and safe mode options (\texttt{options(spam.cholsymmetrycheck= FALSE, spam.safemodevalidity= FALSE)}). The computer code is available as supplementary material.
% In this Section, we use two examples to illustrate the advantages of the chromatic sampler with respect to the two traditional sampling strategies, block and single-site Gibbs sampling. The first example is simulated data arranged according to a regular lattice as encountered in image analysis. The second example considers a Gaussian Markov random field over the irregular lattice formed by the counties of the United States of America.

\subsection{Simulated Imaging Problem}\label{ssec:ImgRest}
Image analysis involves attempting to reconstruct a true latent image, where the `image' may mean a true physical structure as in clinical medical imaging, or an activation pattern or signal as in, e.g., functional magnetic resonance imaging \citep{Lazar08}. The available data consist of pixel values, often corresponding to color on the grayscale taking integer values from 0 to 255. The true values are assumed to have been contaminated with error due to the image acquisition process. This area is one of the original motivating applications for Markov random fields \citep{Besag86, BYM91}.

\begin{figure}[tb]
    \centering
    % sizing = L B R T
    \includegraphics[clip= TRUE, trim= 0in 1.5in 0in 1.5in, scale= 0.5]{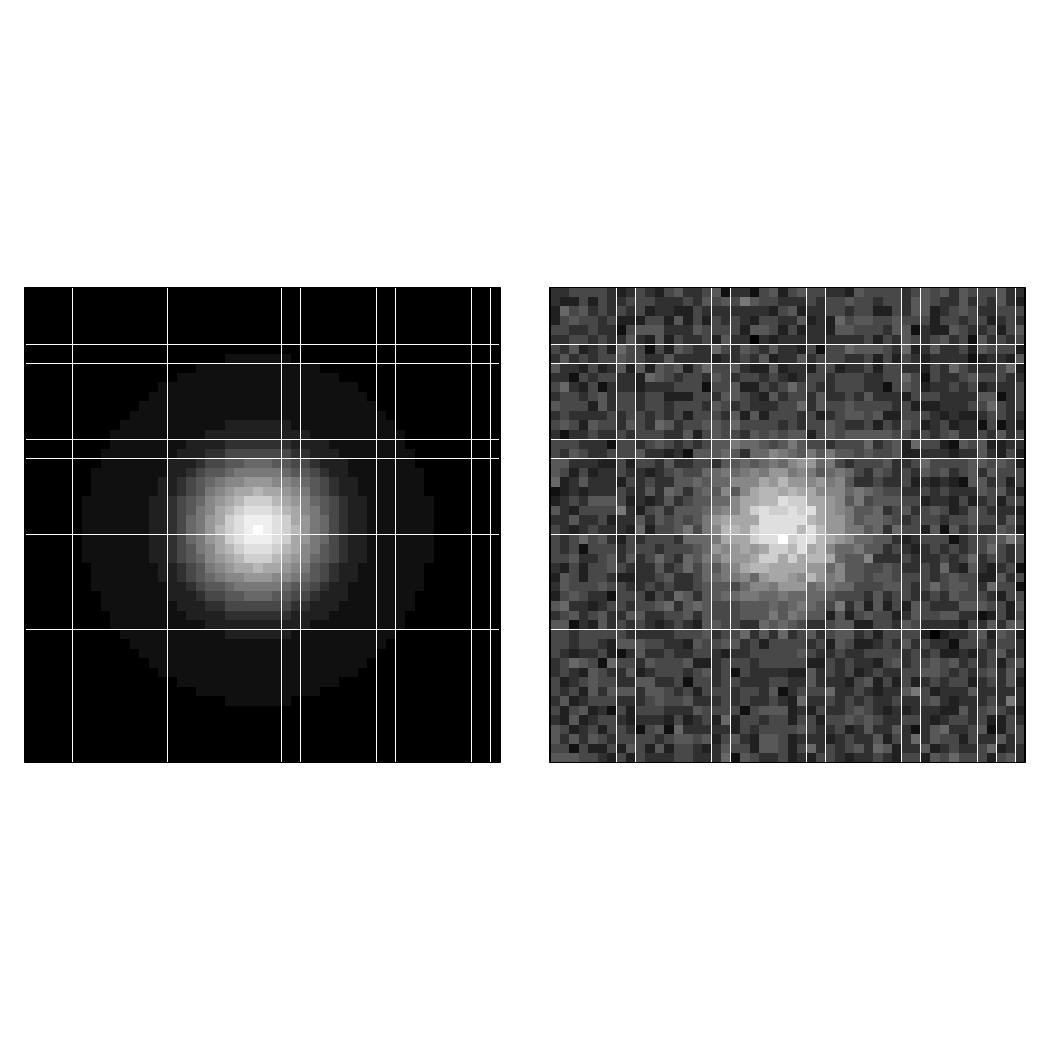}
    \caption{\small{True image (left panel) and corrupted image (right panel) for the simulated image reconstruction example. (These particular images have resolution $50 \times 50$.)}}\label{fig:Data}
\end{figure}

There is growing interest in the statistical analysis of ultra-high dimensional imaging data. For example, structural magnetic resonance images of the human brain may consist of 20-40 two dimensional slices, each of which has $256\times 256$ resolution or higher. Spatial Bayesian models for even a single slice of such data can involve GMRFs over lattices of dimension $n = 256^2$ \citep{BrownMc17} and thus are very computationally challenging when drawing inference via Markov chain Monte Carlo.
Motivated by such applications, we consider images consisting of $p\times p$ pixels, each of which has an observed value $y_{ij}=x_{ij}+\varepsilon_{ij}$, where $x_{ij}$ is the true value of the $(i,j)^{\text{th}}$ pixel in the latent image and $\varepsilon_{ij}$ represents the corresponding contamination. To simulate the data, we take the error terms to be independent, identically distributed $N(0,1)$ random variables. The true image in this case is a rescaled bivariate Gaussian density with $x_{ij}=5 \exp\{ -\|\B{v}_{ij}\|^2/2 \}/\pi$, where $\B{v}_{ij} = (v_i, v_j) \in [-3, 3] \times [-3, 3]$ denotes the center of the $(i,j)^{\text{th}}$ pixel, evenly spaced over the grid, and $\|\cdot\|$ denotes the usual Euclidean norm. Figure \ref{fig:Data} depicts the true generated image (in $50 \times 50$ resolution) and its corrupted counterpart. To study each of the three sampling algorithms, we consider first an image with dimension $n = p \times p = 50^2$.

The assumed model for the observed image is given by $\B{y} = \B{1}\beta_0 + \B{ \gamma} + \B{\varepsilon},$
%\begin{equation}\label{eqn:REMod}
%    \B{y} = \B{1}\beta_0 + \B{ \gamma} + \B{\varepsilon},
%\end{equation}
where $\B{y} \in \mathbb{R}^n$ is the vector of the observed pixel values, $\B{1} = (1, \ldots, 1)^T \in \mathbb{R}^n$, $\beta_0 \in \mathbb{R}$ is a constant intercept parameter, $\B{\gamma}$ is the vector of spatial effects, and $\B{\varepsilon}$ is the vector of errors assumed to follow $N(\B{0}, \sigma^2\B{I})$. To capture local homogeneity of the image, we assume the spatial random effects obey an IAR model with mean zero; i.e., the density of $\B{\gamma}$ is $f(\B{\gamma}) \propto (\tau^2)^{-(n-1)/2}\exp\{ (2\tau^2)^{-1}\B{\gamma}^T(\B{D}-\B{W})\B{\gamma}\}, ~\B{\gamma} \in \mathbb{R}^n$, where $\B{W} = \{w_{ij} := I(i \sim j)\}_{i,j=1}^n$ is the incidence matrix of the underlying graph and $\B{D} = \text{diag}\left(\sum_{j=1}^nw_{ij} : ~i= 1, \ldots, n\right)$. Here we assume a first-order neighborhood structure in which each interior pixel has eight neighbors. We ignore edge effects induced by the perimeter pixels of the image. We specify inverse gamma priors for the variance components and a flat prior for the intercept; i.e., $\sigma^2 \sim \text{InvGam}(\alpha,\alpha)$, $\tau^2 \sim \text{InvGam}(\alpha,\alpha), ~\alpha > 0$, and $\pi(\beta_0) \propto 1, \beta_0 \in \mathbb{R}$. To approximate vague priors for the variance components, we take $\alpha = 0.001$. It has been observed that an inverse gamma prior on $\tau^2$ sometimes can yield undesirable behavior in the posterior \citep{Gelman06}; but our focus is on sampling the random field and thus we use this prior simply for convenience. For posterior sampling, our modeling assumptions lead to a Gibbs sampler having the following full conditional distributions: $\beta_0|\B{y},\B{\gamma},\sigma^2  \sim  N\left( \B{1}^T(\B{y}-\B{\gamma})/n  ,\sigma^2/n\right)$, $\sigma^2|\B{y},\B{\gamma},\beta_0  \sim  \text{InvGam}\left(\alpha+n/2, \alpha+ \|\B{y}-\B{1}\beta_0-\B{\gamma}\|^2/2\right) $, $\tau^2|\B{\gamma}  \sim  \text{InvGam}\left(\alpha+(n-1)/2, \alpha+\B{\gamma}^T(\B{D}-\B{W}) \B{\gamma}/2\right)$, and $\B{\gamma}|\B{y},\sigma^2,\tau^2   \sim  N\left( \B{Q}_p^{-1}\B{b},\B{Q}_p^{-1}\right)$,
%\begin{eqnarray*}
%    \beta_0|\B{y},\B{\gamma},\sigma^2 & \sim & N\left( \B{1}^T(\B{y}-\B{\gamma})/n  %,\sigma^2/n\right) \\
%    \sigma^2|\B{y},\B{\gamma},\beta_0 & \sim & IG\left(\alpha+n/2, \alpha+ %\|\B{y}-\B{1}\beta_0-\B{\gamma}\|^2/2\right) \\
%    \tau^2|\B{\gamma} & \sim & IG\left(\alpha+n/2, \alpha+\B{\gamma}^T(\B{D}-\B{W}) %\B{\gamma}/2\right) \\
%    \B{\gamma}|\B{y},\sigma^2,\tau^2  & \sim & N\left( \B{Q}^{-1}\B{b},\B{Q}^{-1}\right),
%\end{eqnarray*}
where $\B{Q}_p=\sigma^{-2}\B{I} + \tau^{-2}(\B{D}-\B{W})$ and $\B{b}= (\B{y}-\B{1}\beta_0)/\sigma^2$. {We remark that it is not unusual for spatial prior distributions to have zero or constant mean functions \citep{BayarriEtAl07}, since the {\em a posteriori} updated spatial model will still usually capture the salient features. In the presence of noisy data, however, identifiability is limited, meaning that the parameters will more closely follow the assumed correlation structure in the prior. In terms of statistical efficiency, this situation favors block sampling over componentwise updating, as previously mentioned.}

To implement the Gibbs sampler, three strategies are employed, with the only difference being how we sample the full conditional distribution of $\B{\gamma}$. First, since $\B{Q}_p$ is sparse, we consider full block Gibbs sampling based on Algorithm 1 in the Supplementary Material to sample $\B{\gamma}$ in a single block. The second strategy is to obviate the large matrix manipulation by employing single-site Gibbs sampling using the local characteristics, $\gamma_i|\B{\gamma}_{(-i)},\B{y},\sigma^2,\tau^2  \sim  N(\mu_i, \sigma^2_i), ~i= 1, \ldots, n$, where $\mu_i  =  \sigma^2_i (\sigma^{-2} y_i + \tau^{-2} \sum_{j \in \mathcal{N}(i)}w_{ij} \gamma_{j}  ) $ and $\sigma^2_i  =  \tau^2\sigma^2 \{\sigma^2(\B{D})_{ii} +\tau^2\}^{-1}$.
%\begin{eqnarray}
%    \begin{aligned}
%        \mu_i  &=  \frac{\tau^2 y_i + \sigma^2 \sum_{j \in \mathcal{N}(i)}w_{ij} \gamma_{j}    %}{\sigma^2(\B{D})_{ii} +\tau^2}\\
%        \sigma^2_i  &=  \frac{\tau^2  \sigma^2 }{\sigma^2(\B{D})_{ii} +\tau^2},\\
%    \end{aligned}\label{eqn:SSG}
%\end{eqnarray}
The final sampling strategy we implement is chromatic Gibbs sampling discussed in Subsection \ref{ssec:HGS}. This approach uses the coloring depicted in Figure \ref{fig:colEx} as a 4-coloring of the pixels in the image. Following the notation in Subsection \ref{ssec:HGS}, we have that $\gamma_i \mid \B{\gamma}_{C_j} ,\B{y},\sigma^2,\tau^2 \stackrel{\text{indep.}}{\sim} N(\mu_i, \sigma_i^2), ~i \in \mathcal{A}_j, ~j=1,...,4.$ The most important feature of the chromatic sampler is that $\gamma_i \mid \B{\gamma}_{C_j} ,\B{y},\sigma^2,\tau^2, ~~i \in \mathcal{A}_j,$ can be drawn simultaneously. All of the necessary conditional means and variances for a given color can be computed through {matrix-vector multiplication. In general, multiplcation of an $n\times n$ matrix with an $n\times 1$ vector has $\mathcal{O}(n^2)$ complexity. Similar to the block sampler, though, we use sparse representations of the necessary matrices for the chromatic sampler, which reduces the complexity to $\mathcal{O}(n)$ since each pixel has relatively few Markov neighbors.}

We implement the three sampling strategies so that each procedure performs $10,000$ iterations of the Monte Carlo Markov chain to approximate the posterior distribution of the model parameters. For each approach, three chains are run using dispersed initial values. We assess convergence of the chains via trace plots, Gelman plots \citep{BrooksGelman98}, and plots of cumulative ergodic averages of scalar hyperparameters. We discard the first 8000 iterations as a burn-in period and assess convergence using the last 2000 realizations of each Markov chain. The simulations, coded entirely in \verb|R|, are carried out on a Dell Precision T3620 desktop running Windows 10 with an Intel Xeon 4.10 GHz CPU and 64 GB of RAM.

\begin{figure}[tb]
    \centering
    % sizing = L B R T
    \includegraphics[clip= TRUE, trim= 0in 0in 0in 0in, scale= 0.45]{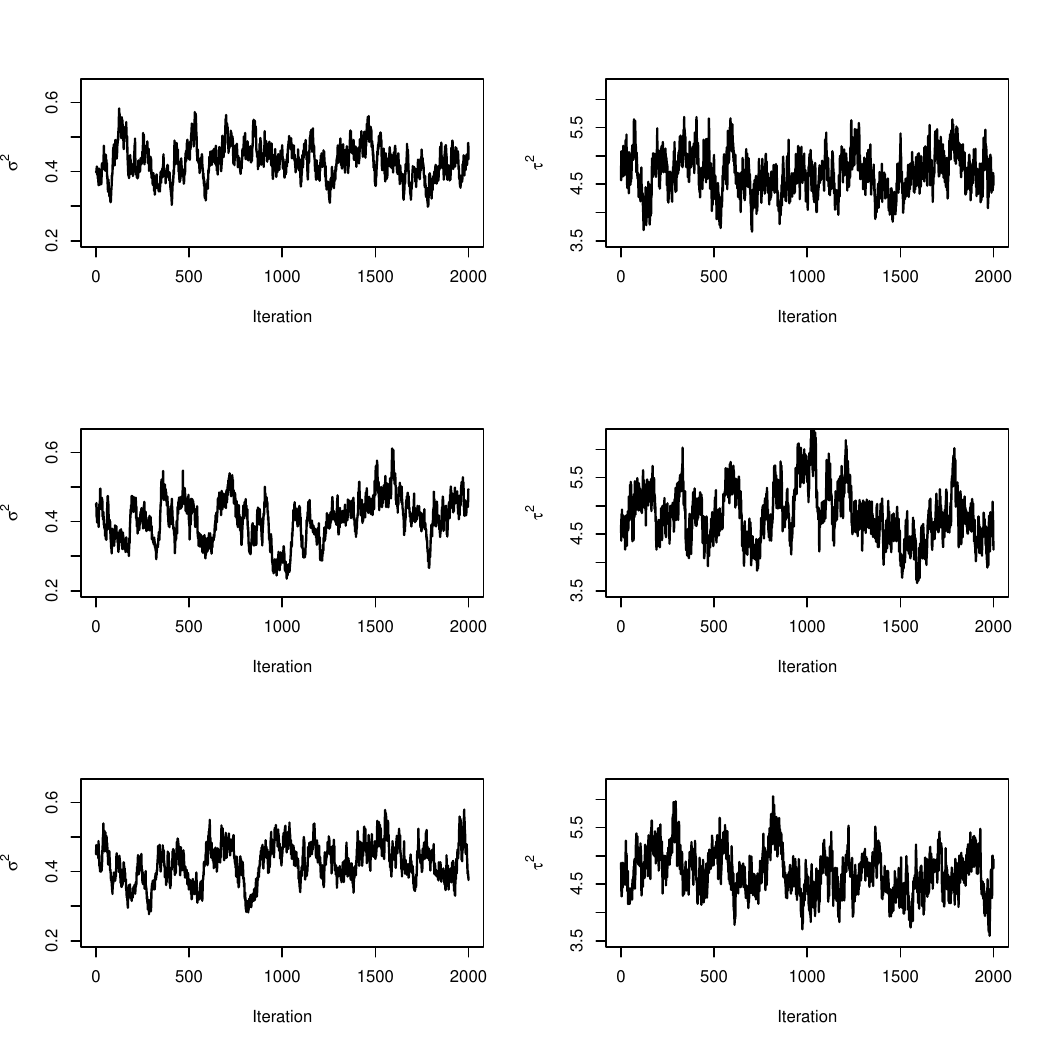}
    \includegraphics[clip= TRUE, trim= 0in 0in 0in 0in, scale= 0.45]{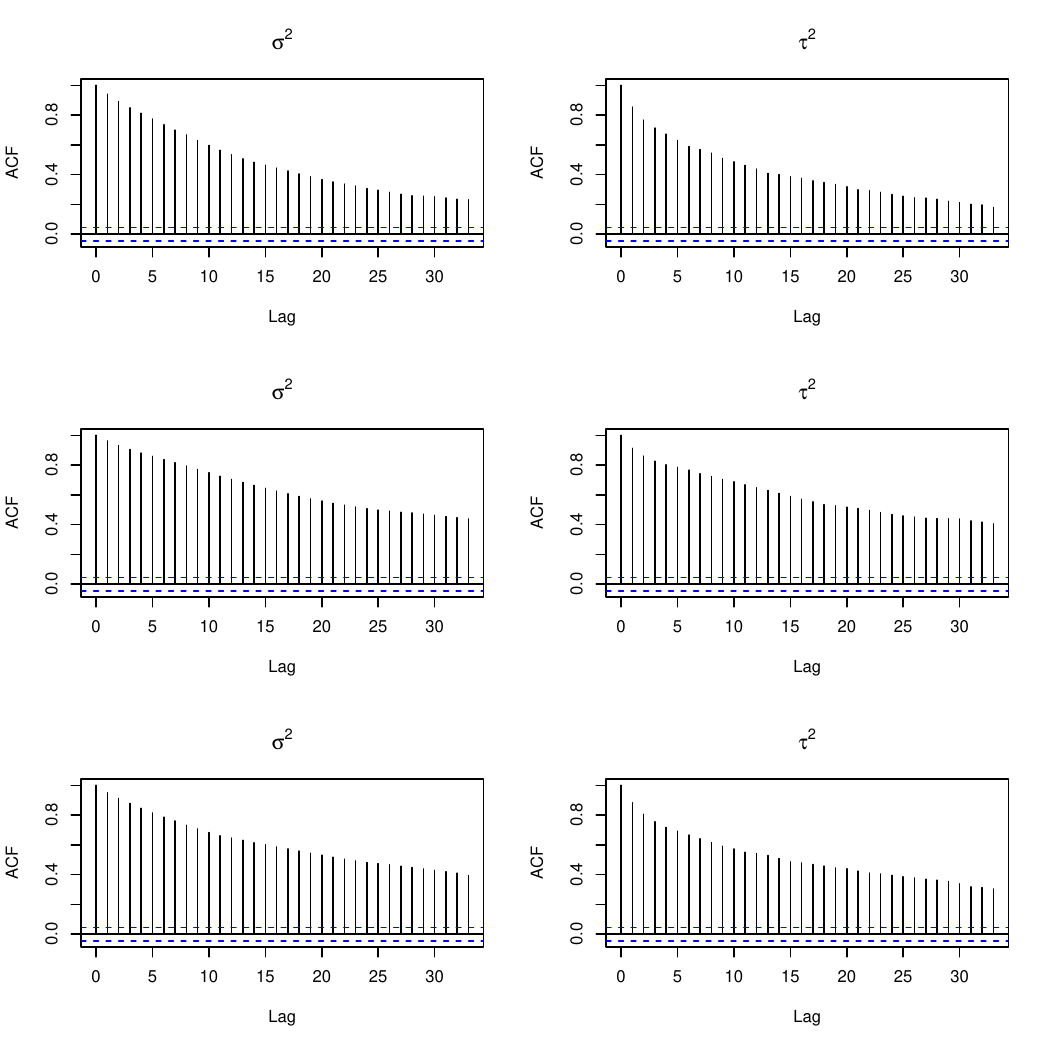}
    \caption{\small{MCMC Trace plots (two left columns) and empirical ACF plots (two right columns) of single chains each for $\sigma^2$ and $\tau^2$ for the $50\times 50$ regular array example. The top, middle, and bottom rows are from the chromatic, block, and single-site chains, respectively.}}\label{fig:lnTrace}
\end{figure}

\begin{figure}[tb]
    \centering
    % sizing = L B R T
    \includegraphics[clip= TRUE, trim= 0in 0in 0in 0in, scale= 0.35]{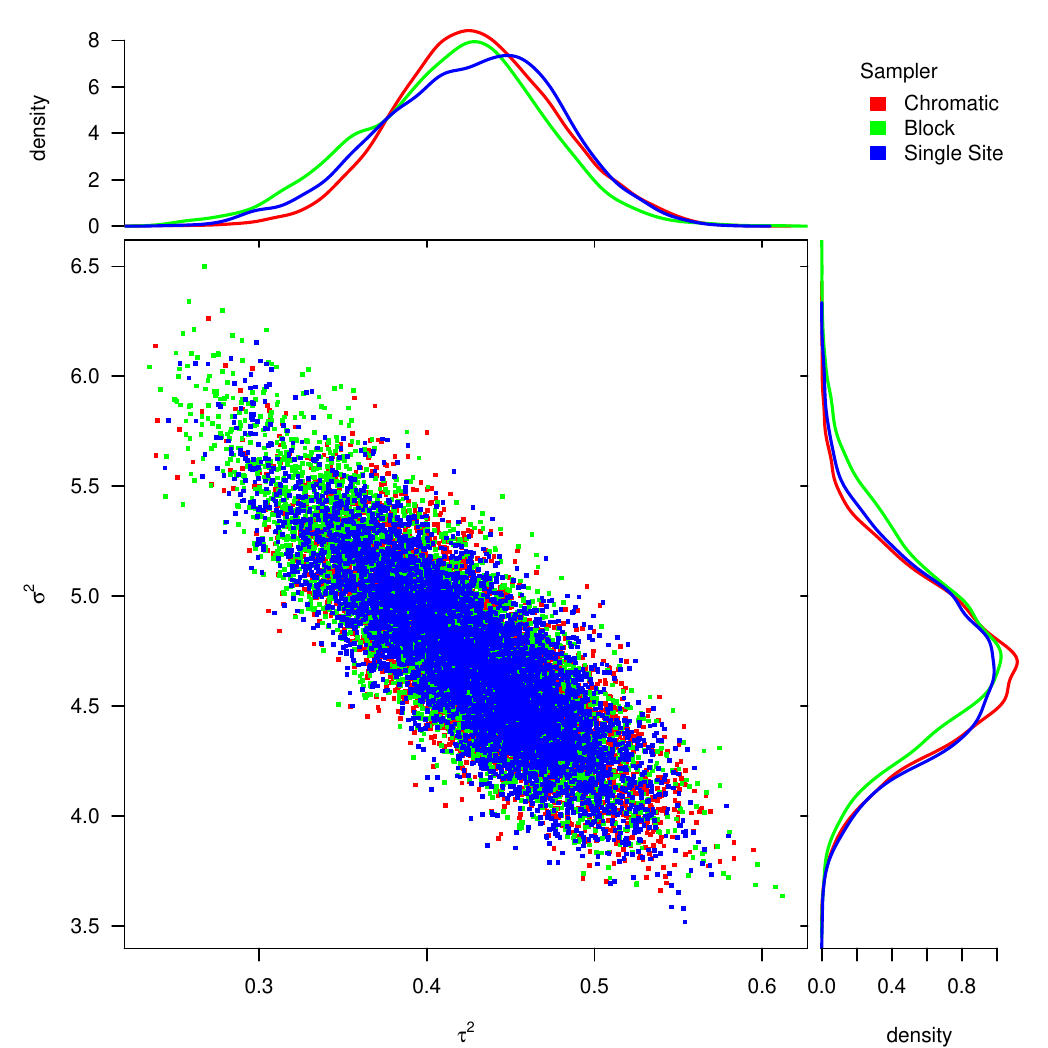}
    \includegraphics[clip= TRUE, trim= 0in 0in 0in 0in, scale= 0.35]{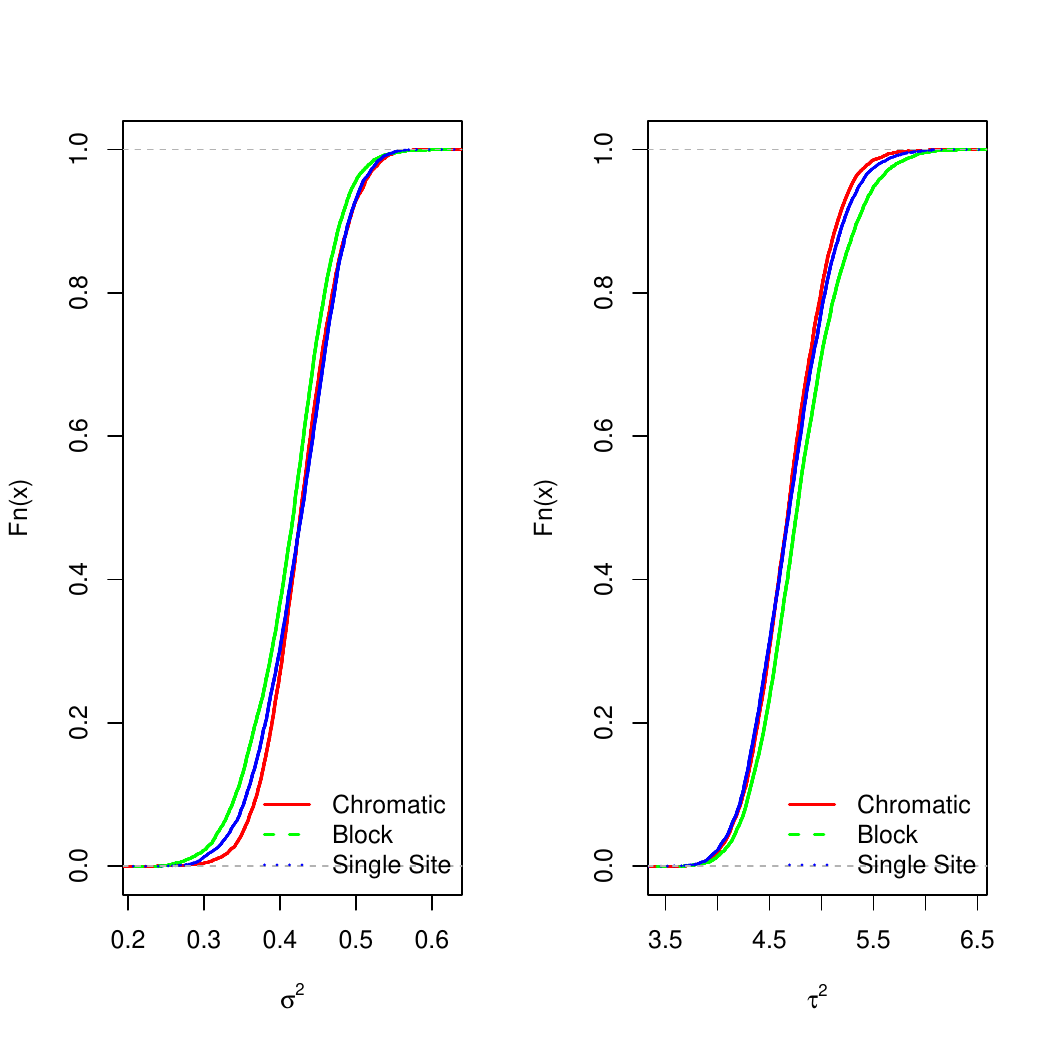}
    \caption{\small{Left panel: Scatterplot and estimated marginal posterior densities (left) and empirical CDFs (right) from the three sampling approaches in the $50\times 50$ array example. The left panel was created using code available at \url{https://github.com/ChrKoenig/R_marginal_plot}.}}\label{fig:lnDists}
\end{figure}

\begin{figure}[!tb]
    \centering
    % sizing = L B R T
    \includegraphics[clip= TRUE, trim= 0in 0.5in 0in 1in, scale= 0.3]{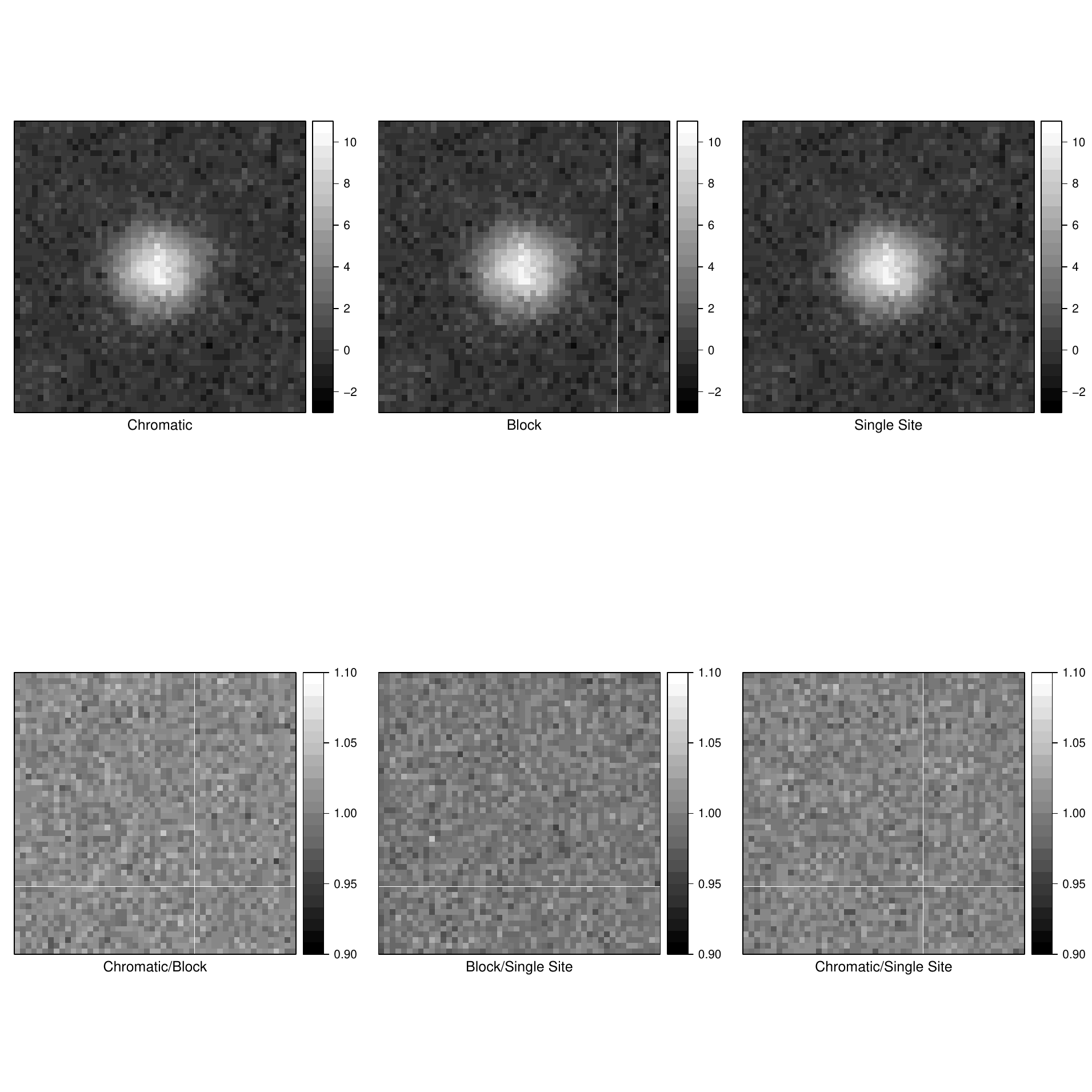}
    \caption{\small{Posterior mean estimates of the true image obtained from each sampling approach (top row) along with pairwise standard deviation ratios (bottom row) in the $50\times 50$ array example.}}\label{fig:lnMeans}
\end{figure}

Figure \ref{fig:lnTrace} displays the trace plots and empirical autocorrelation functions for the hyperparameters $\sigma^2$ and $\tau^2$ for chromatic, block, and single-site sampling. We see very similar behavior in terms of autocorrelation across all three sampling approaches. From Supplementary Figure 1, we glean that each sampling approach has approximately converged in the $\sigma^2$ and $\tau^2$ chains after 2,000 iterations, although the block sampler evidently has the longest convergence time according to the Gelman plots. While each approach produces estimates of $\sigma^2$ and $\tau^2$ that tend to the same value, the block sampler exhibits slightly larger Monte Carlo standard error than the other two approaches. This partly explains the slight difference in empirical distribution from the block sampler versus that of the chromatic and single-site samplers, as depicted in Figure \ref{fig:lnDists}. Regardless, the joint and marginal density estimates largely agree. This agreement is also evident in Figure \ref{fig:lnMeans}, which displays the posterior mean estimates of the true image $\beta_0\B{1} + \B{\gamma}$, the primary quantity of interest. To assess exploration of the posterior distributions, the Figure also depicts point-wise ratios of sample standard deviations for each pair of algorithms. All three samplers produce distributional estimates that are virtually indistinguishable.

% \begin{figure}[tb]
%     \centering
%     % sizing = L B R T
%     \includegraphics[clip= TRUE, trim= 0in 0in 0in 0in, scale= 0.45]{5050LowNoiseGelman.pdf}
%     \includegraphics[clip= TRUE, trim= 0in 0in 0in 0in, scale= 0.45]{5050LowNoiseCumAvg.pdf}
%     \caption{\small{Left panel: Gelman plots \citep{BrooksGelman98} of the potential scale reduction factors versus chain length for the $50\times 50$ regular array example. Right panel: Cumulative averages $\hat{\sigma^2}^{(k)}$ and $\hat{\tau^2}^{(k)}, ~k= 1, \ldots, 10,000$ calculated from three independent chains. In the right panel, the top, middle, and bottom rows correspond to chromatic, block, and single-site sampling, respectively.}}\label{fig:lnCumPlot}
% \end{figure}

The primary advantage of the chromatic approach versus the other two is in the computational cost incurred to obtain each sample. Of course, the same number of samples from two different algorithms is not guaranteed to provide the same quality of approximation to the target distribution. To accommodate the different convergence characteristics of the three algorithms while still considering total computation time (including the burn-in period), we measure the {\em cost per effective sample} \citep{FoxNorton16}, $CES := N^{-1}\kappa T$, where $T$ is the total computation time, $N$ is the size of the retained sample from the Markov chain (after burn-in), and $\kappa$ is the integrated autocorrelation time \citep[IAT;][]{KassEtAl98, CarlinLouis09}. $CES$ measures the total computational effort required to generate an effectively independent sample from the target distribution. Table \ref{tab:5050Costs} displays the total CPU times, effective sample sizes, integrated autocorrelation times, and $CES$ for the $\tau^2$ chains under each sampling approach. Here we see an approximately {89\%} improvement in computational effort between independent samples compared to block Gibbs sampling. Single-site sampling (when coded in \texttt{R}) is by far the worst performer, as expected. It is interesting to note that in this case, the chromatic sampler has the shortest IAT of the three methods considered.
\begin{table}[tb]
    \caption{\small{CPU times to draw 2,000 realizations (including 8,000 burn-in iterations) from one $\tau^2$ Markov chain under each sampling approach in the $50\times 50$ array example. Also reported are the effective sample sizes (ESS), integrated autocorrelation times (IAT), and costs per effective sample (CES).}}
    \centering
    {\small
    \begin{tabular}{l | c c c c}
        Sampler & CPU Time ($s$) & ESS & IAT & CES\\
        \hline
        Chromatic & 10.99 & 65.74 & 30.42 & 0.17 \\
        Block & 49.63 & 32.71 & 61.15 & 1.52 \\
        Single-Site & 3331.68 & 53.54 & 37.36 & 62.23 \\
    \end{tabular}
    }
    \label{tab:5050Costs}
\end{table}

% Increasing dimension with noisy data
To further study the performance of block sampling versus chromatic sampling, particularly how they scale with regular arrays of increasing dimension, we repeat the model fitting procedure using data simulated as before, but with images of size $p \times p$, for $p = 80, 128, 256,$ {and $512$}. To create a more challenging situation, we add considerably more noise to the images by assuming $Var(\B{\varepsilon}) = 50^2\B{I}$. This makes the underlying spatial field much more weakly identified by the data and thus more strongly determined by the prior. Hence the GMRF parameters ($\beta_0, \B{\gamma}$, $\tau^2$) will be more strongly correlated in the posterior, creating a more challenging situation for any MCMC algorithm. For each $p$, we run the same model with the same prior specifications as in the first example. We again run each MCMC algorithm for 10,000 iterations, treating the first 8,000 as burn-in periods.

\begin{figure}[tb]
    \centering
    % sizing = L B R T
    \includegraphics[clip= TRUE, trim= 0in 1.75in 0in 1.5in, scale= 0.6]{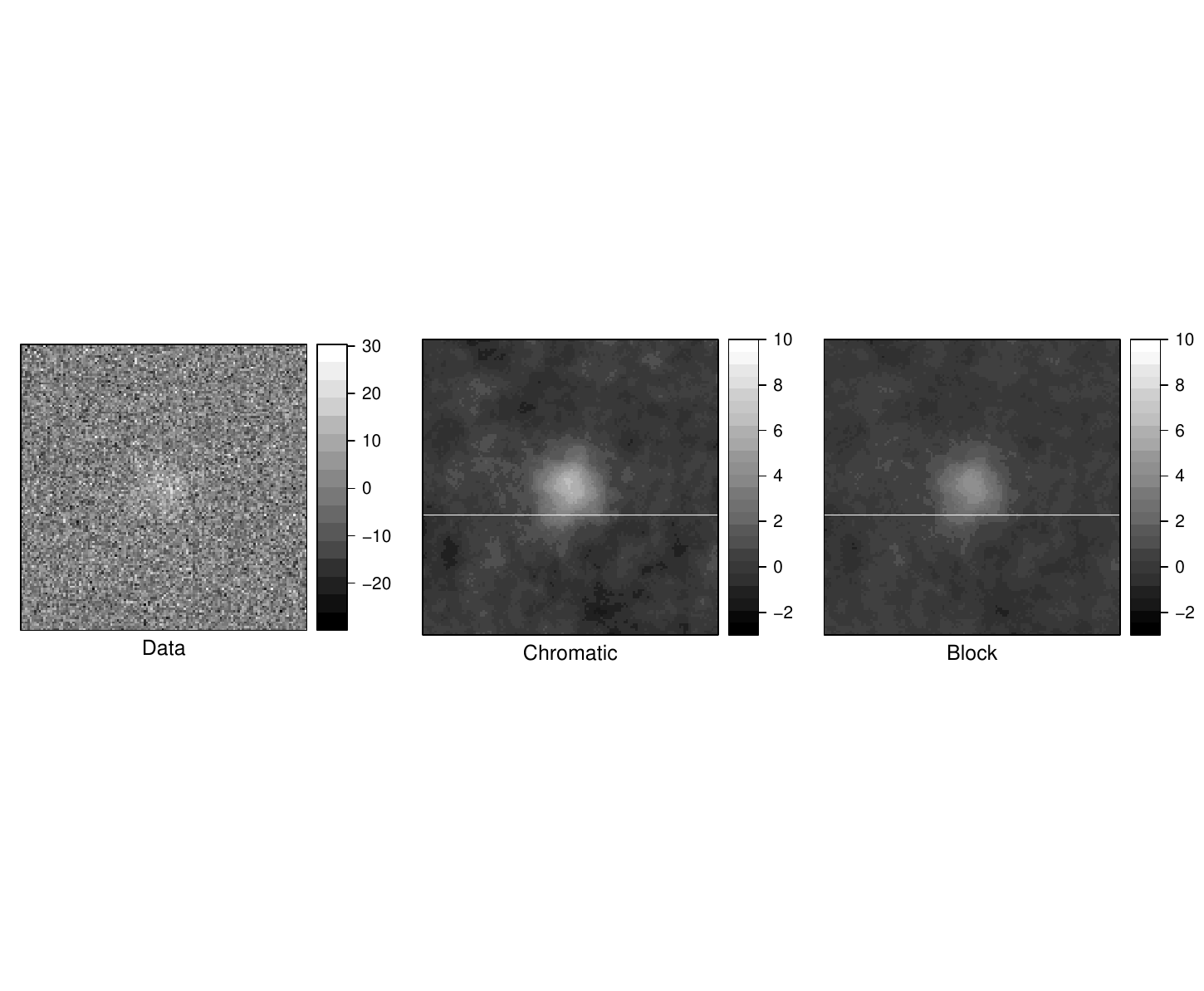}\\
    \caption{\small{Simulated data and posterior mean estimates of the true image from the chromatic and block sampling approaches in the noisy $128\times 128$ array example.}}\label{fig:hn128Means}
\end{figure}

Supplementary Figures 2 through 8 display diagnostics and posterior mean estimates produced by the different sampling procedures under $p = 50, 80, 128$ with noisy data. The \texttt{R}-coded single-site sampler was not computationally feasible for images with resolution $p \geq 80$ and so was not considered. As expected, we see the autocorrelation in the $\tau^2$ chains increased with the noisy data, regardless of the sampling approach, whereas the data-level variance $\sigma^2$ remains well identified. The three approaches still produce parameter estimates that agree with each other. {The Gelman plots indicate that the chromatic sampler takes longer to converge than the other two approaches, but still becomes an acceptable approximation to a posterior sample after about $7,000$ iterations. The joint and marginal densities of $(\sigma^2, \tau^2)$ are more diffuse than the low-noise case, again agreeing with intuition.} Despite the deterioration of the $\tau^2$ chain, the quantity of interest as in many imaging problems is a function of the model parameters. In this case, we are mainly interested in $\B{\varphi} := \beta_0\B{1} + \B{\gamma}$, meaning that we want to explore the so-called {\em embedded posterior distribution} of $\B{\varphi}$. The parameter $\B{\varphi}$ (i.e., the underlying image) converges well under chromatic and block sampling. Hence we are able to recover a reasonable approximation of the target image, as evident in Figure \ref{fig:hn128Means} and Supplementary Figure 7. This phenomenon echoes the observation of \cite{GelfandSahu99} that even when a Gibbs sampler is run over a posterior distribution that includes poorly identified parameters ($\tau^2$ in this case), inferences can still be drawn for certain estimands living in lower dimensional space than the full posterior.

% CPU / Memory requirements for increasing dimension
{As noted at the beginning of this Section, the preceding results are obtained by only coding in \texttt{R} and without any parallel processing. However, if one is interested in accelerating the conditional GMRF updating step, they might choose to simply code the single-site sampler in \texttt{C} or \texttt{Fortran} and incorporate it into a larger MCMC algorithm. On the other hand, a researcher might want to fully exploit the ability to do chromatic updates in parallel, as opposed to vectorized updates. To examine the computational gains obtained by a simpler algorithm in a faster language, we run the single-site algorithm with the simulated imaging data, but where the GMRF updating step is passed to \texttt{C++} via the \texttt{Rcpp} package \citep{EddRcpp11}. Further, we implement the chromatic sampler with truly parallelized updates in \texttt{R} by distributing the independent updates (those corresponding to the same color in the graph) over eight processors (five in the $50 \times 50$ case)  via the \texttt{parallel} package \citep{R16}. In terms of the generated Markov chains, the \texttt{Rcpp} single site sampler and parallel chromatic sampler are algorithmically identical to the \texttt{R}-coded single-site and vectorized chromatic, respectively, so we do not look at their convergence characteristics separately. The code for implementing the \texttt{C++} and parallel approaches is also available as supplementary material.}

\begin{figure}[tb]
    \centering
    % sizing = L B R T
    \includegraphics[clip= TRUE, trim= 0in 0in 0in 0in, scale= 0.45]{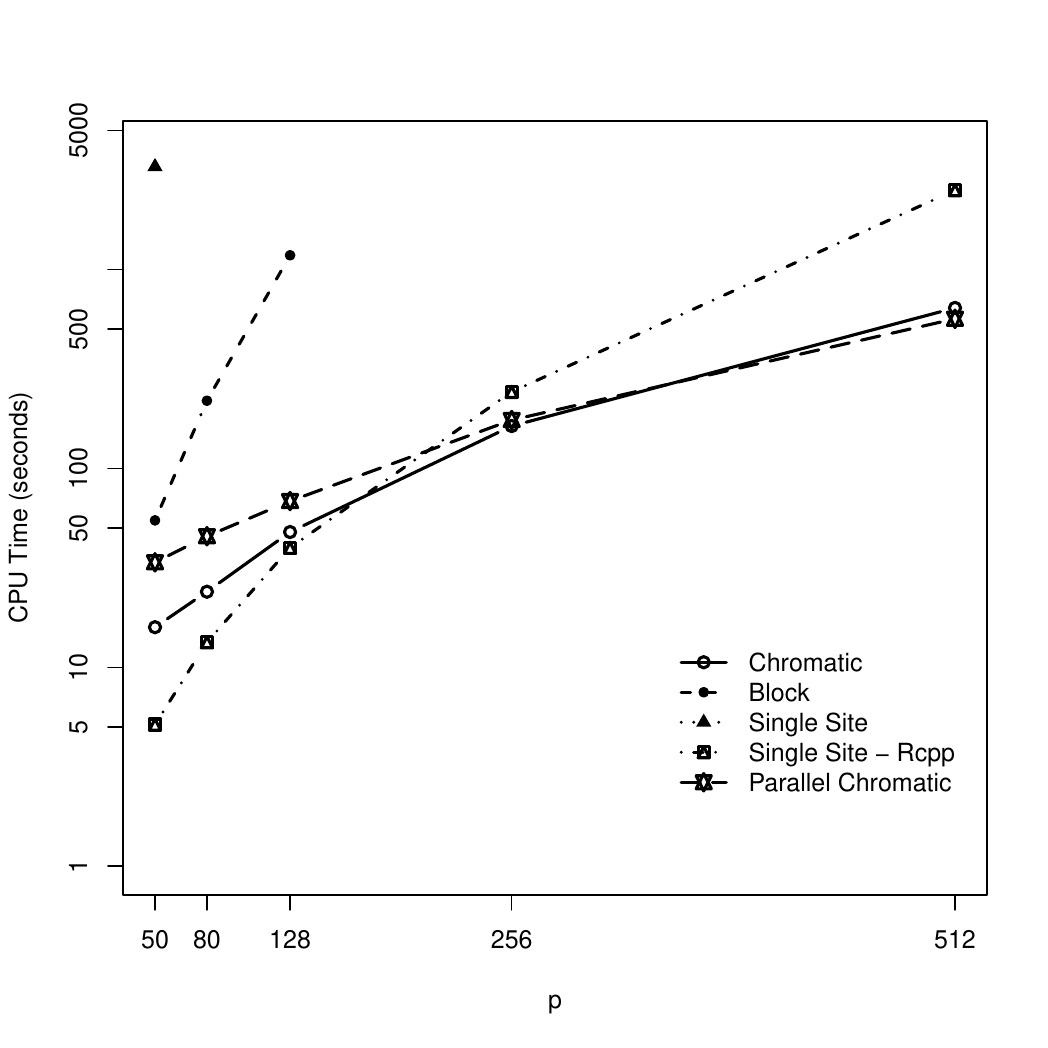}
    \includegraphics[clip= TRUE, trim= 0in 0in 0in 0in, scale= 0.45]{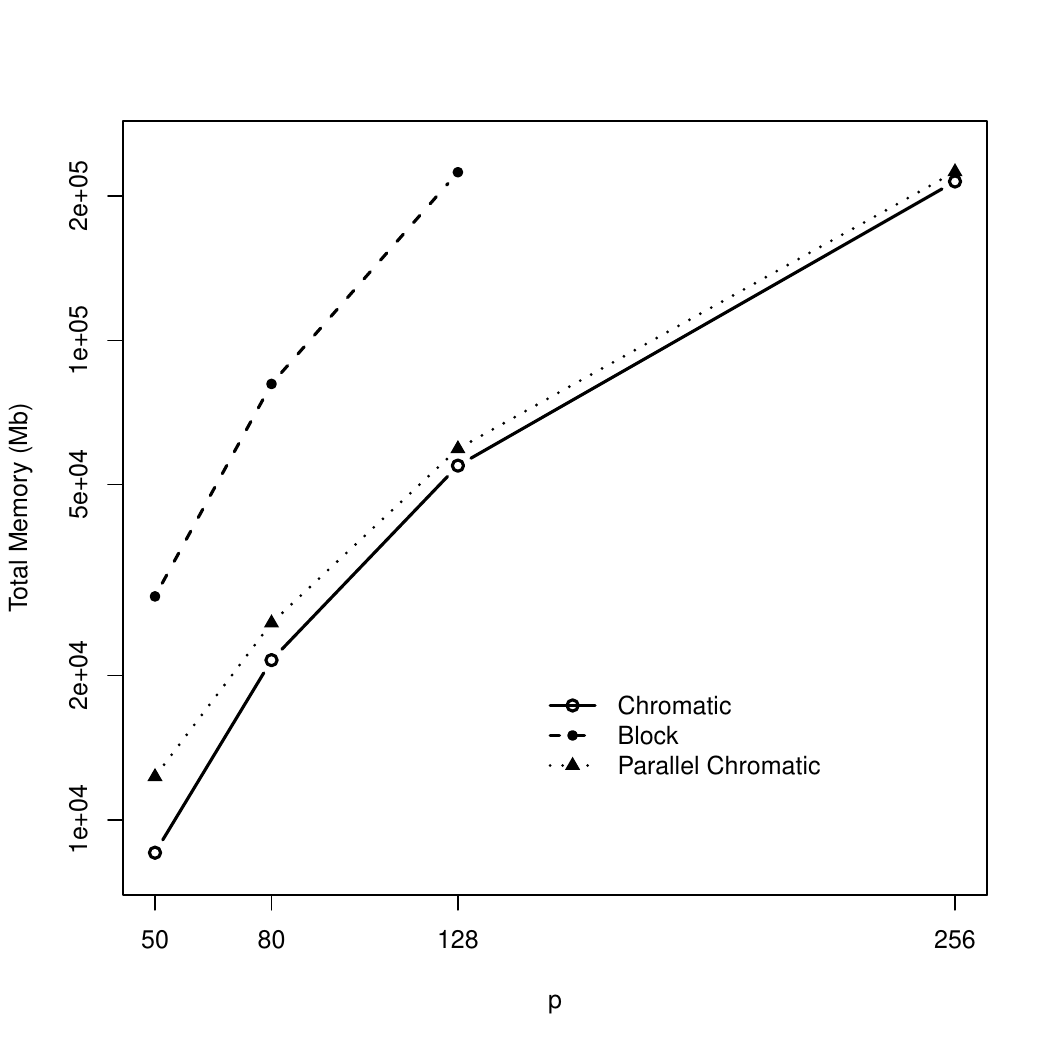}
    \caption{\small{CPU time (left) and total memory required (right) for the different sampling implementations to complete 10,000 iterations for simulated noisy $p\times p$ arrays. In both plots, the $y$ axis is on the log scale. (Note that memory is only tracked up to $p=256$, and single site memory usage was not tracked.)}}\label{fig:hnCPUMem}
\end{figure}

Figure \ref{fig:hnCPUMem} displays the total CPU time required to complete 10,000 iterations for $p= 50, 80, 128, 256, 512$. As previously mentioned, the {\texttt{R}-coded} single-site sampler is only feasible in the $p = 50$ case, as it is by far the most inefficient implementation due to the nested loops. Chromatic sampling requires much less computing time than block sampling, and scales at a lower rate This is due in part to the fact that no Cholesky factorizations are required for chromatic sampling. Such factorizations with even sparse matrices can be expensive, and repeated multivariate Gaussian draws are still required even when the symbolic factorization is stored throughout the MCMC routine. {The parallel implementation of the chromatic sampler requires more CPU time at lower resolutions, but is more scalable than the vectorized version. The cost of parallelization becomes comparable to the vectorized implementation at $256\times 256$ and is slightly faster than the vectorization at $p = 512$. This illustrates how the overhead associated with distributing data across processors cancels out any computational gain at smaller scales. Parallelizing becomes worthwhile for extremely large datasets in which splitting up a huge number of pixels is worth spending the overhead. There is also overhead associated with passing the data and parameter values to a function written in \texttt{C++}, as we see in the \texttt{Rcpp} implementation of the single-site sampler. At small to moderate resolutions, the \texttt{Rcpp} version is much faster than the chromatic sampler and the block sampler. However, the computational cost of \texttt{Rcpp} scales at a much faster rate than the chromatic versions, so much so that both chromatic versions are an order of magnitude faster than the \texttt{C++} single-site sampler at $p = 512$. Similar to how a parallel implementation depends on the size of the data, we see that the benefit of coding a chromatic sampler in \texttt{R} versus passing to \texttt{C++} is more pronounced when the dataset is extremely large.}

There is also considerable memory overhead associated with both sparse Cholseky block updating and chromatic sampling. The total required memory for Cholesky-based block updating depends on the storage scheme used by the sparse matrix implementation. The \verb|spam| implementation in our example uses a variant of the so-called compressed sparse row format \citep{Sherman75}. Chromatic sampling, on the other hand, requires no matrix storage at all, but only lists of identifiers associated with each graph color. The right panel of Figure \ref{fig:hnCPUMem} illustrates the consequent savings in total memory allocations and how they scale with arrays of increasing dimension. {In terms of total memory allocations, both the vectorized and parallel chromatic samplers require considerably less than block updating.} In fact, block updating for $p = 256$ and $p = 512$ was not possible due to memory limitations. The Cholesky factorization failed, returning \verb|Cholmod error: `problem too large'|.

\subsection{Binomial Election Data on an Irregular Lattice}

Here we examine the performance of the block Gibbs and chromatic sampling strategies on an irregular lattice, since both the structure of $\B{Q}$ and the possible colorings of the underlying graph are more complicated. Moreover, we illustrate the performance of these procedures when applied to non-Gaussian data. In particular, we examine geographical trends in voter preference using binomial outcomes. The data were obtained from the Harvard Election Data Archive (\url{https://dataverse.harvard.edu}) and are depicted in Supplementary Figure 10.

Our data consist of polling results from the 2010 New York Governor's race in which Democratic candidate Andrew Cuomo defeated Republican candidate Carl Paladino and Green Party candidate Howie Hawkins. During this election, the state of New York had $14,926$ precincts, with polling data being available on $14,597$ precincts. The 329 precincts for which data are unavailable is attributable to improper reporting or lack of voter turnout.

Let $Y_i$ be the number of votes cast for the Democratic candidate out of $m_i$ total votes in precinct $i, ~i =1, \ldots, n$. Then we assume that $Y_i | \pi_i, m_i \stackrel{\text{indep.}}{\sim} \text{Bin}(m_i,\pi_i)$, where $g^{-1}(\pi_i) = \beta_0 + \gamma_i$, $g(\cdot)$ is the usual logistic link function, and $\boldsymbol{\gamma} = (\gamma_1,..., \gamma_n)^T$ is a vector of random effects inducing spatial homogeneity. We suppose that $\B{\gamma}$ follows a {\em proper} IAR model; i.e., $\B{\gamma} \sim N(\B{0}, \tau^2(\B{D} - \rho\B{W})^{-1})$, where $\B{D}$ and $\B{W}$ are as defined in Section \ref{ssec:ImgRest}. Here, the ``propriety parameter" $\rho \in (\lambda_1^{-1}, \lambda_{n}^{-1})$ ensures that the precision matrix is non-singular, where $\lambda_1 < 0$ and $\lambda_N > 0$ are the smallest and largest eigenvalues of $\B{D}^{-1/2}\B{W}\B{D}^{-1/2}$, respectively \citep{BanerjeeEtAl15}. Proper IARs are sometimes used as approximations to the standard IAR when a proper prior distribution is desired. For simplicity, we fix $\rho = 0.995$. The model is completed with the prior assumptions that $\beta_0\sim N(0,1000)$ and $\tau^2 \sim \text{InvGam}(1,1)$.
% In this analysis, each precinct will be viewed as a unique spatial location. Thus, the available data consists of the number of votes cast for the Democratic candidate at the $i$th precinct ($Y_i$) and the total number of votes cast ($m_i$), for $i=1,...,n$. For modeling purposes, it is assumed that $Y_i | p_i, m_i \sim \text{Binomial}(m_i,p_i)$, where $g^{-1}(p_i) = \beta_0 + \gamma_i$, $g(\cdot)$ is the usual logistic link, and $\boldsymbol{\gamma} = (\gamma_1,..., \gamma_n)^T$ is a vector of spatial random effects which obey a CAR model. In particular, the considered CAR model takes the form
% $ N(\boldsymbol{0},\tau^2(\boldsymbol{D} - \rho \boldsymbol{W})^{-1})$, where $\boldsymbol{D}$ and $\boldsymbol{W}$ are defined as in section \ref{ssec:ImgRest} with the convention that two precincts are said to be neighbors if they share a common border. Note, the only difference in the CAR model considered here and the IAR model from \ref{ssec:ImgRest} is the propriety parameter $\rho$, with the addition of $\rho$ supplying prior propriety. \cite{BanerjeeEtAl15} advocates for the use of the IAR for the reason of prior interpretability, citing that posterior propriety is generally attained. Thus, to closely emulate the IAR model but guarantee prior propriety, $\rho$ is taken to be near unity in this application; i.e., $\rho=0.995$.

Under the logistic link, we can simplify posterior sampling via data augmentation. This technique exploits the fact that $\exp(\eta)^a(1 + \exp(\eta))^{-b} = 2^{-b} \exp(\kappa \eta) \int_0^\infty \exp(- \psi \eta^2/2) p(\psi|b,0)$ $d \psi$, where $\eta \in \mathbb{R}$, $a \in  \mathbb{R}$, $b \in \mathbb{R}^+$, $\kappa = a - b/2$, and $p(\cdot|b,0)$ is the probability density function of a P\'{o}lya-Gamma random variable with parameters $b$ and $0$ \citep{Polson13}. Using this identity, the observed data likelihood can be written as $\pi(\mathbf{Y}|\beta_0, \boldsymbol{\gamma}, ) \propto   \prod_{i=1}^n \exp\{\kappa_{i}\eta_{i}\}$ $\times\int_0^\infty \exp(- \psi_{i} \eta_{i}^2/2) p(\psi_{i}|m_{i},0) d \psi_{i}$,
where $\boldsymbol Y=(Y_1,...,Y_n)^T$, $\eta_i=\beta_0 + \gamma_i$ and $\kappa_{i}=Y_{i}-m_{i}/2$. Thus, by introducing $\psi_{i}$ as latent random variables, we have that  $\pi(\boldsymbol{Y}, \boldsymbol \psi|\beta_0, \boldsymbol{\gamma})  \propto \exp \{ - \left( (\beta_0\B{1} + \B{\gamma})^T \boldsymbol D_{\boldsymbol \psi} (\beta_0\B{1} + \B{\gamma}) - 2\boldsymbol \kappa^T \boldsymbol (\beta_0\B{1} + \B{\gamma}) \right)/2 \}\prod_{i =1}^n p(\psi_{i}|m_{i},0)$,
where $\boldsymbol \psi=(\psi_1,...,\psi_n)^T$, $\boldsymbol \kappa= (\kappa_1,...,\kappa_n)^T$, and $\boldsymbol D_{\boldsymbol \psi}=\textrm{diag}(\boldsymbol \psi)$. By including $\B{\psi}$ in the MCMC algorithm, we induce a Gaussian full conditional distribution on $\B{\gamma}$, facilitating GMRF updates without having to tune a Metropolis-Hastings algorithm. Additional implementation details are provided in the Supplementary Material.

% Under these specifications, a posterior sampling algorithm can be constructed under which Gibbs steps are used to update all parameters of the model. For further details, see the Supplementary Material. The salient point, the form of the full conditional distribution of $\boldsymbol\gamma$ is such that the chromatic Gibbs sampler, full block Gibbs, and single-site sampler can be used. That said, the number of spatial regions is large enough that the single-site sampler is for all intents and practical purposes computationally infeasible to apply.

In order to implement chromatic sampling, a coloring of the underlying Markov graph has to be found. Using the greedy algorithm given in Algorithm 2 in the Supplementary Material, we obtain a 7-coloring, so that the chromatic sampler can update the entire $n = 14,926 $-dimensional field in seven steps.
The coloring is depicted in Supplementary Figure 11.

\begin{figure}[tb]
    \centering
    \begin{tabular}{c}
    \includegraphics[clip= TRUE, trim= 1in 1in 1in 1in, scale= 0.35]{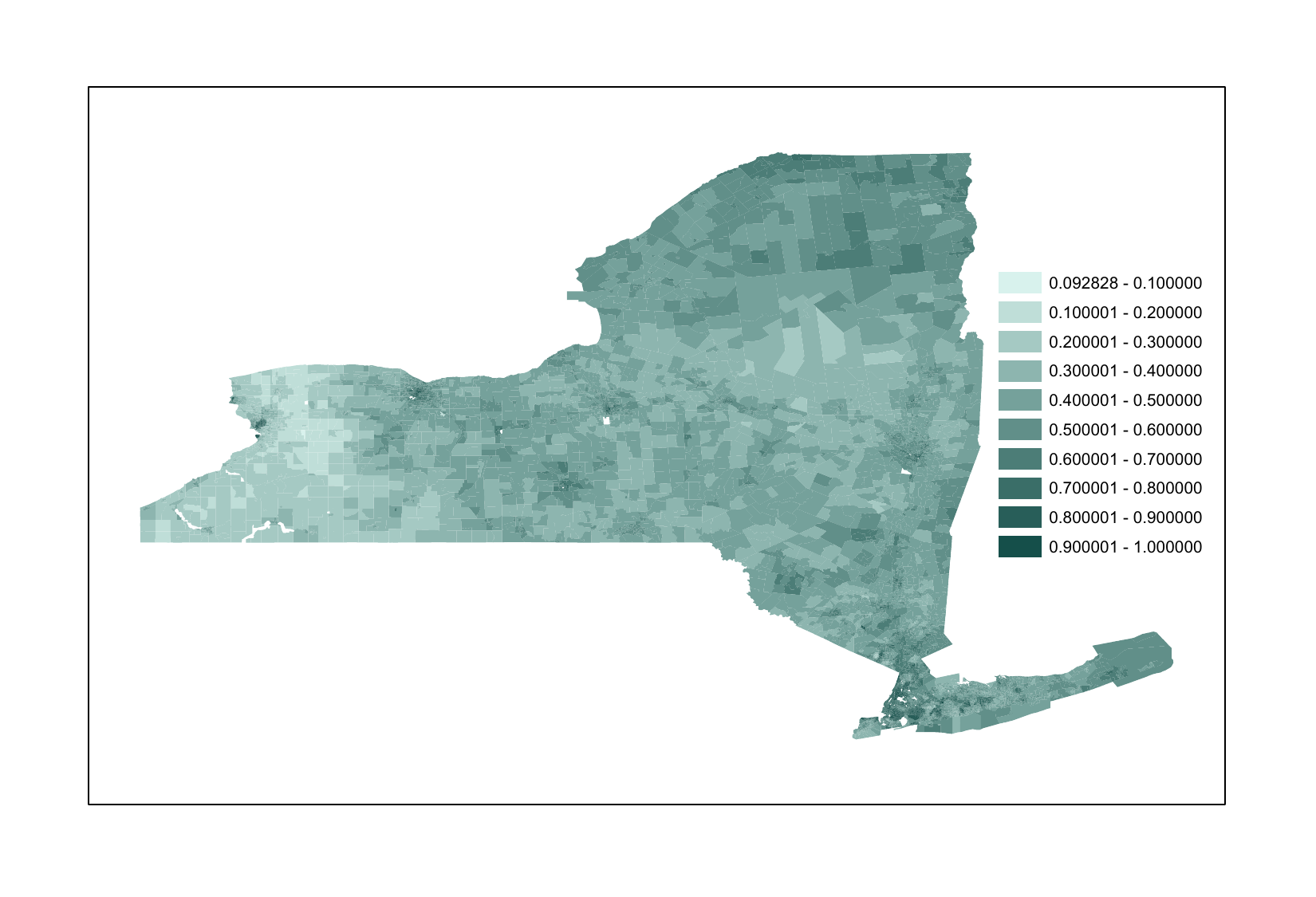} \\
    \includegraphics[clip= TRUE, trim= 1in 1in 1in 1in, scale= 0.35]{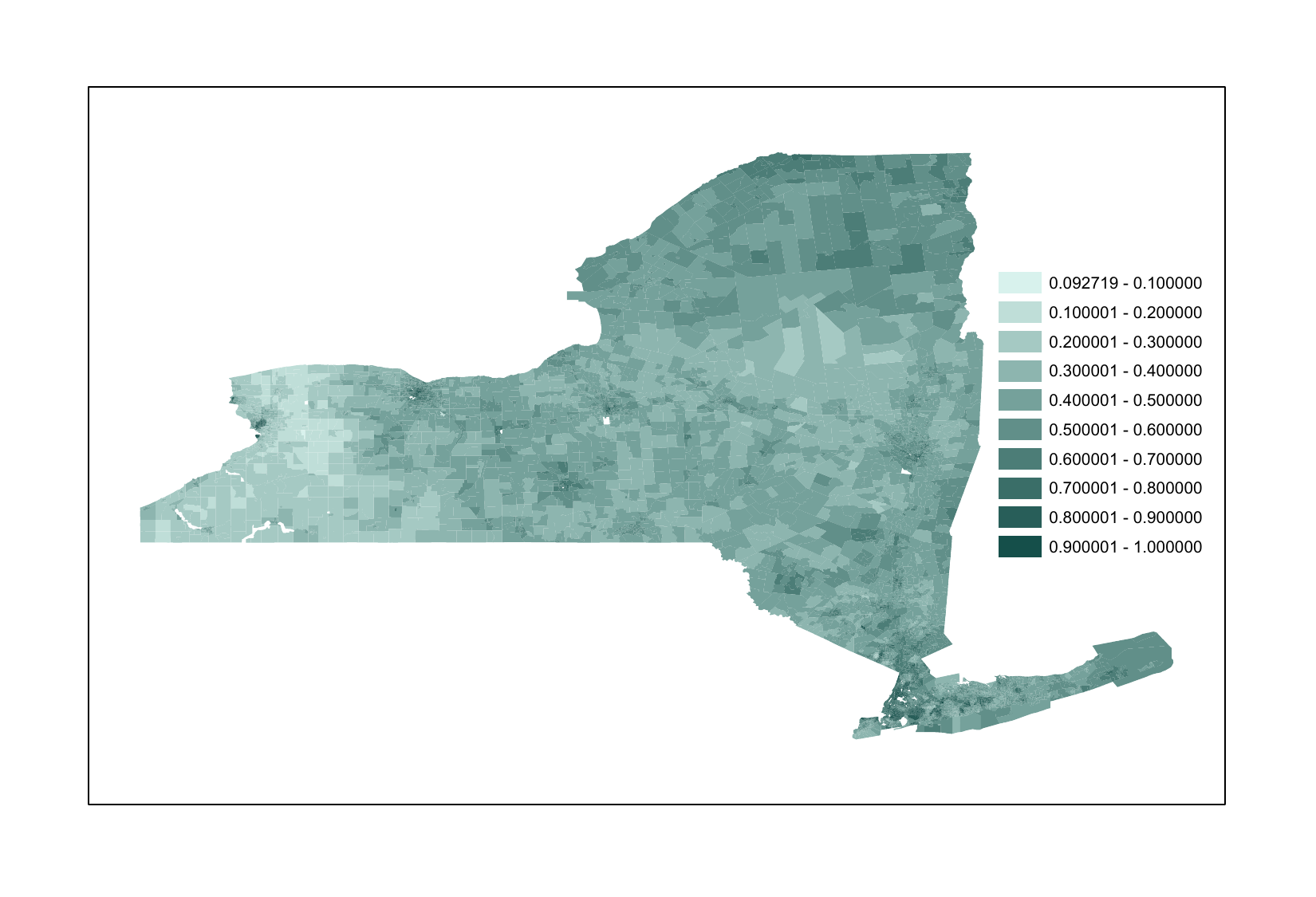}
    \end{tabular}
    \caption{\small{Posterior mean maps of voter preference for the Democratic candidate in the binomial election example obtained from the full block Gibbs (top), and chromatic Gibbs (bottom) sampling.}}\label{fig:Pest}
\end{figure}

We implement Gibbs sampling with both the block Gibbs and chromatic updates for $10,000$ iterations, discarding the first $5,000$ as a burn-in period. The code is run on a desktop using Windows 10 with an Intel Core i5-3570 3.40GHz CPU with 16GB of RAM. The trace plots and empirical ACF plots for $\beta_0$ and $\tau^2$ are depicted in Supplementary Figure 12, along with the Gelman plots of these two parameters in Supplementary Figure 13. We see adequate convergence in the same number of iterations under both sampling approaches. Table \ref{tab:Pest} summarizes the results for both samplers in terms of CPU time and cost per effective sample of the intercept and variance terms. We again see a savings in CPU time under chromatic sampling, so much so that it offsets the slightly larger autocorrelation time. Thus we are able to obtain effectively independent samples with less computational effort. The posterior mean maps of the voter Democratic preference ($\pi_i$) obtained under each sampling strategy are displayed in Figure \ref{fig:Pest}. We see essentially identical results under both strategies.
% These results reinforce the findings from our simulation described in Subsection \ref{ssec:ImgRest}. As before, the CBG sampler is far more computationally efficient when compared to the FBG, even if the data are arranged according to an irregular lattice.

\begin{table}[tb]
    \caption{\small{CPU times to draw 5,000 realizations (after 5,000 burin-in iterations) from one Markov chain under each sampling approach in the New York election example. Also reported are the effective sample sizes (ESS), autocorrelation times (ACT), and costs per effective sample (CES).}}
    \centering
    \begin{tabular}{l | c c c c}
        Sampler & CPU Time (s) & ESS & ACT & CES \\ \hline
         $\beta_0$ Chromatic & 222.06& 2445.74 &2.04 &0.0935 \\
         $\beta_0$ Block &  294.16 & 2732.82 &1.83 &0.1018\\
        $\tau^2$ Chromatic & 222.06& 1916.63 &2.61 & 0.1155\\
         $\tau^2$ Block &  294.16 & 2186.021 &2.29 &0.1332
    \end{tabular}
     \label{tab:Pest}
\end{table}

\subsection{Summary}
These numerical experiments illustrate potential improvements that chromatic Gibbs sampling can offer versus the two most common strategies of block sampling and single-site sampling. In simulated image reconstruction, we find that for every considered resolution, the chromatic sampler is computationally much cheaper than the full block Gibbs and single-site samplers coded entirely in \texttt{R}. We observe in both chromatic and block sampling the deterioration in Monte Carlo Markov chains that is known to occur as the dimension of a GMRF increases \citep{RueHeld05, agapiou2014analysis}. Even in this case, however, any of the approaches considered are able to estimate the posterior mean of the latent field and obtain equivalent recovery of the quantity of interest. The chromatic sampler is able to do so much more quickly and with much less memory overhead, the latter of which allows the chromatic sampler to scale to images of extremely large dimension beyond the capability of standard Cholesky factorization routines available in \verb|R|. The potential advantages extend to irregular arrays and non-Gaussian data, as demonstrated in the election data example.

% The cost per effective sample is also consistently smaller, suggesting sufficient convergence to the target distribution with much less computational effort. Moreover, as the dimension of the problem grows, the gains in computational efficiency become more pronounced. The chromatic Gibbs sampler completes the sampling process roughly 4, 6, 10, and 15 times faster than even the full block Gibbs sampler for $p= 25, ~50, ~75, ~100$, respectively. The single-site sampler is the worst performer, often taking more than 100 times longer than chromatic Gibbs.
%We achieve similar gains when the data are on an irregular lattice, even with a coloring that is not guaranteed to be optimal. Once we can find a coloring with, say, a greedy algorithm, we are in a position to dramatically accelerate MCMC computations involving GMRFs.

\section{DISCUSSION}\label{sec:disc}
Over the last twenty years, Gaussian Markov random fields have seen a dramatic increase in popularity in the applied Bayesian community. In this work, we discussed approaches for simulating from Gaussian Markov random fields that are commonly used in practice. We compared the two dominant approaches in the statistics literature, single-site and block updating, to chromatic Gibbs sampling. Each procedure has theoretical guarantees, but our criteria have been pragmatic; i.e., how can statisticians effectively lower the computational cost of sampling from the target distribution without resorting to esoteric knowledge from graph theory, numerical analysis, or parallel programming? Taking this view, we have shown that chromatic sampling is competitive with and often able to improve upon single-site and full block Gibbs. {In a large-scale scenario, we demonstrated improvements afforded by chromatic sampling even when compared to passing the single-site updating step to \texttt{C++}. This shows that computational efficiency gains are achievable even when a researcher desires (or needs) to do as much programming as possible in a high-level interpreted language.}

% In summary, given the current trajectory of modern data analysis, the utility of GMRFs is not likely to diminish anytime soon. However, with their use comes the need for efficient yet accessible sampling strategies to facilitate Bayesian posterior inference in large-scale problems. Given the findings presented herein, we believe that the chromatic sampler can be used to meet that need.

Motivated by large-scale clinical imaging data, we illustrated potential advantages on a regular array with Gaussian response, finding that chromatic sampling scales to settings where memory limitations prevent direct sparse matrix manipulations. We also considered a real example with binomial election data on an irregular lattice with almost 15,000 areal units, showing that chromatic sampling is useful even without a provably optimal coloring of the MRF graph. Both block sampling and chromatic sampling tend to be far superior to single-site sampling when working in \verb|R|. %Further, the relative computational savings of chromatic sampling versus block sampling can be dramatic as the dimension of the problem grows, even in the presence of sparse precision matrices.

While facilitating parallel or vectorized simultaneous updates, each individual draw under chromatic sampling is still at the level of a single site. Thus, for variables that are highly correlated in the target distribution, convergence can be slow. To handle this, \citet{GonzalezEtAl11} propose also a ``splash sampler" to combine the blocking principle of updating sets of correlated variables together with the parallelizability afforded by graph colorings. Splash sampling is more involved than simple chromatic sampling. It requires careful construction of undirected acyclic graphs subject to a known tree width determined by individual processor limitations, and hence much more computing effort and more familiarity with graph theory. In the Gaussian case, splash sampling would require repeated Cholesky factorizations, each on matrices of smaller dimension, but without being able to save the sparsity structure. In the presence of highly correlated variables in the target distribution with GMRF updates, it might be preferable to use ordinary block updates with sparse matrix algebra and the algorithms suggested by \cite{Rue01}. However, our numerical experiments demonstrate that the gain in computational efficiency from the simple chromatic sampler can still outweigh the loss of statistical efficiency. This leads to an overall improvement in a variety of situations without resorting to more sophisticated approaches that might be inaccessible to most statisticians.

{In both the \texttt{C++} and parallel implementations, we assume that the researcher would prefer to keep as much of the code in \texttt{R} as possible due to its user-friendly functionality and their desire to avoid low-level programming. Since \texttt{R} itself is a software suite written primarily in \texttt{C}, even matrix-vector calculations and vectorized functions are ultimately executed via loops in \texttt{C} (or a another low-level language such as \texttt{Fortran}). Thus, if one were to code our entire MCMC algorithm with `vectorized' chromatic updates in \texttt{C} or \texttt{C++}, the result would be an algorithm that is essentially identical to a single-site sampler up to the order in which the sites are updated. In other words, if a researcher is working purely in \texttt{C} or \texttt{Fortran}, then the computational advantages of chromatic sampling can only be fully realized with a distributed parallel approach. However, parallel programming in such low-level languages is much more difficult and nuanced than it is in \texttt{R} and thus beyond the expertise of many statisticians and data scientists. Indeed, one of the main reasons for the popularity of \texttt{R} is the ease with which extremely complicated tasks (e.g., MCMC on exotic hierarchical Bayesian models or sparse matrix factorizations) can be executed. There is nothing \texttt{R} can do that cannot be done in a low-level language if one has the time and patience to write the code for it. Our purpose in this work is to discuss and demonstrate computational savings that may be realized without having to leave the more comfortable environment of a high-level, interpreted programming language.}

{The parallel implementation used in this work used only eight processors, but this still showed modest acceleration over vectorized chromatic sampling with the largest dataset we considered. The difference would no doubt be much more pronounced by distributing the effort over more processors. With the advent of modern computing clusters and GPU computing, it is becoming more common for researchers to have available thousands of cores for use simultaneously. In fact, it is not difficult to envision scenarios where the number of processors available is $\mathcal{O}(n)$, in which case the complexity of parallel chromatic sampling reduces to $\mathcal{O}(k)$, where $k$ is often fixed as $n$ increases due to the structure of the data \citep{GonzalezEtAl11}. Thus the scaling potential of parallel chromatic sampling is enormous and worthy of further investigation. We defer such an exploration to future work.}

In this paper, we examined the performance of chromatic sampling versus single-site and block Gibbs on high density data in which the entire study region is sufficiently sampled and in which a first-order Markov neighborhood can capture the salient features of the data. This situation is applicable to many, but not all, analyses of areal data. There remains the issue of how chromatic sampling would perform in the presence of sparse observations from an underlying smooth process, where the autocorrelation of the Markov chain would be expected to be higher than in the high-density case. {Related to this point is that, to the best our knowledge, the convergence rates associated with chromatic sampling Markov transition kernels remain unknown. We leave these questions to be explored at a later date.}

Given the current trajectory of modern data analysis, the utility of GMRFs is not likely to diminish anytime soon. However, with their use comes the need for efficient yet accessible sampling strategies to facilitate Bayesian posterior inference along with appropriate measures of uncertainty. This area remains an active area of research among statisticians, computer scientists, and applied mathematicians. Fortunately, the increasingly interdisciplinary environment within which researchers are operating today makes it more likely that significant advancements will be widely disseminated and understood by researchers from a wide variety of backgrounds. This is no doubt a promising trend which will ultimately benefit the broader scientific community.\\

%\section*{Supplementary Material}
%\bigskip
%\begin{center}
%{\large\bf SUPPLEMENTARY MATERIAL}
%\end{center}
%
%\begin{description}
%
%\item[TASsupp:] This Supplementary Material contains additional figures referred to in the text and details concerning the posterior sampling algorithm used in the election example. (.pdf file)
%
%\item[RCode:] The relevant \texttt{R} code used for implementing the numerical examples (compressed .zip file)
%
%\end{description}

%\section*{Acknowledgements}
%
%This material is based upon work partially supported by the National Science Foundation (NSF) under Grant DMS-1127914 to the Statistical and Applied Mathematical Sciences Institute. DAB is partially supported by NSF Grants CMMI-1563435 and EEC-1744497. CSM is partially supported by National Institutes of Health Grant R01 AI121351.

\bibliographystyle{asa}
\bibliography{HGSref}

\begin{thebibliography}{63}
\newcommand{\enquote}[1]{``#1''}
\expandafter\ifx\csname natexlab\endcsname\relax\def\natexlab#1{#1}\fi

\bibitem[{Agapiou et~al.(2014)Agapiou, Bardsley, Papaspiliopoulos, and
  Stuart}]{agapiou2014analysis}
Agapiou, S., Bardsley, J.~M., Papaspiliopoulos, O., and Stuart, A.~M. (2014),
  \enquote{Analysis of the {G}ibbs sampler for hierarchical inverse problems,}
  \textit{SIAM/ASA Journal on Uncertainty Quantification}, 2, 511--544.

\bibitem[{Banerjee et~al.(2015)Banerjee, Carlin, and Gelfand}]{BanerjeeEtAl15}
Banerjee, S., Carlin, B.~P., and Gelfand, A.~E. (2015), \textit{{Hierarchical
  Modeling and Analysis for Spatial Data}}, Boca Raton: Chapman \& Hall/CRC,
  2nd ed.

\bibitem[{Bardsley(2012)}]{Bardsley12}
Bardsley, J.~M. (2012), \enquote{{MCMC-based image reconstruction with
  uncertainty quantification},} \textit{SIAM Journal on Scientific Computing},
  34, 1316--1332.

\bibitem[{Bates and Maechler(2016)}]{BatesMaechlerMatPackage}
Bates, D. and Maechler, M. (2016), \textit{{Matrix: Sparse and dense matrix
  classes and methods}}, {R package version 1.2-6}.

\bibitem[{Bayarri et~al.(2007)Bayarri, Berger, Paulo, Sacks, Cafeo, Cavendish,
  Lin, and Tu}]{BayarriEtAl07}
Bayarri, M.~J., Berger, J.~O., Paulo, R., Sacks, J., Cafeo, J.~A., Cavendish,
  J., Lin, C.-H., and Tu, J. (2007), \enquote{{A framework for validation of
  computer models},} \textit{Technometrics}, 49, 138--154.

\bibitem[{Besag(1974)}]{Besag74}
Besag, J. (1974), \enquote{{Spatial interaction and the statistical analysis of
  lattice systems},} \textit{Journal of the Royal Statistical Society, Series
  B}, 36, 192--236.

\bibitem[{Besag(1986)}]{Besag86}
--- (1986), \enquote{{On the statistical analysis of dirty pictures},}
  \textit{Journal of the Royal Statistical Society, Series B}, 48, 259--302.

\bibitem[{Besag and Kooperberg(1995)}]{BesagKoop95}
Besag, J. and Kooperberg, C. (1995), \enquote{{On conditional and intrinsic
  autogregressions},} \textit{Biometrika}, 82, 733--46.

\bibitem[{Besag et~al.(1991)Besag, York, and Molli\'{e}}]{BYM91}
Besag, J., York, J.~C., and Molli\'{e}, A. (1991), \enquote{{Bayesian image
  restoration, with two applications in spatial statistics},} \textit{Annals of
  the Institute of Statistical Mathematics}, 43, 1--59.

\bibitem[{Br\'{e}laz(1979)}]{Brelaz79}
Br\'{e}laz, D. (1979), \enquote{{New methods to color the vertices of a
  graph},} \textit{Communications of the ACM}, 22, 251--256.

\bibitem[{Brooks and Gelman(1998)}]{BrooksGelman98}
Brooks, S.~P. and Gelman, A. (1998), \enquote{General methods for monitoring
  convergence of iterative simulations,} \textit{Journal of Computational and
  Graphical Statistics}, 7, 434--455.

\bibitem[{Brown et~al.(2017{\natexlab{a}})Brown, Datta, and
  Lazar}]{BrownEtAl17}
Brown, D.~A., Datta, G.~S., and Lazar, N.~A. (2017{\natexlab{a}}), \enquote{{A
  Bayesian generalized CAR model for correlated signal detection},}
  \textit{Statistica Sinica}, 27, 1125--1153.

\bibitem[{Brown et~al.(2014)Brown, Lazar, Datta, Jang, and
  McDowell}]{BrownEtAl14}
Brown, D.~A., Lazar, N.~A., Datta, G.~S., Jang, W., and McDowell, J.~E. (2014),
  \enquote{{Incorporating spatial dependence into Bayesian multiple testing of
  statistical parametric maps in functional neuroimaging},}
  \textit{NeuroImage}, 84, 97--112.

\bibitem[{Brown et~al.(2017{\natexlab{b}})Brown, McMahan, Shinohara, and
  Linn}]{BrownMc17}
Brown, D.~A., McMahan, C.~S., Shinohara, R.~T., and Linn, K.~L.
  (2017{\natexlab{b}}), \enquote{{Bayesian spatial binary regression for label
  fusion in structural neuroimaging},} {ArXiv 1710.10351}.

\bibitem[{Cai(2014)}]{Cai14}
Cai, Z. (2014), \enquote{{Very large scale Bayesian machine learning},}
  Unpublished doctoral dissertation, Rice University, Department of Computer
  Science.

\bibitem[{Cai et~al.(2013)Cai, Jermaine, Vagena, Logothetis, and
  Perez}]{CaiEtAl13}
Cai, Z., Jermaine, C., Vagena, Z., Logothetis, D., and Perez, L. (2013),
  \enquote{{The pairwise Gaussian random field for high-dimensional data
  imputation},} \textit{IEEE 13th International Conference on Data Mining
  (ICDM)}, 61--70.

\bibitem[{Carlin and Louis(2009)}]{CarlinLouis09}
Carlin, B.~P. and Louis, T.~A. (2009), \textit{{Bayesian Methods for Data
  Analysis}}, Boca Raton: Chapman \& Hall/CRC, 3rd ed.

\bibitem[{Cheng et~al.(2015)Cheng, Cheng, Liu, Peng, and Teng}]{ChengEtAl15}
Cheng, D., Cheng, Y., Liu, Y., Peng, R., and Teng, S.-H. (2015),
  \enquote{{Efficient sampling for Gaussian graphical models via spectral
  sparsification},} in \textit{Journal of Machine Learning Research:
  Proceedings of the 28th International Conference on Learning Theory}, pp.
  364--390.

\bibitem[{Culberson(1992)}]{Culberson92}
Culberson, J.~C. (1992), \enquote{{Iterated greedy graph coloring and the
  difficulty landscape},} Technical Report, University of Alberta.

\bibitem[{Dempster(1972)}]{Dempster72}
Dempster, A.~P. (1972), \enquote{{Covariance selection},} \textit{Biometrics},
  28, 157--175.

\bibitem[{Eddelbuettel and Fran\c{c}ois(2011)}]{EddRcpp11}
Eddelbuettel, D. and Fran\c{c}ois, R. (2011), \enquote{{Rcpp}: Seamless {R} and
  {C++} Integration,} \textit{Journal of Statistical Software}, 40, 1--18.

\bibitem[{Fox and Norton(2016)}]{FoxNorton16}
Fox, C. and Norton, R.~A. (2016), \enquote{{Fast sampling in a linear-Gaussian
  inverse problem},} \textit{SIAM/ASA Journal on Uncertainty Quantification},
  4, 1191--1218.

\bibitem[{Furrer and Sain(2010)}]{FurrerSainSpam}
Furrer, R. and Sain, S.~R. (2010), \enquote{{spam: A sparse matrix R package
  with emphasis on MCMC methods for Gaussian Markov random fields},}
  \textit{Journal of Statistical Software}, 36, 1--25.

\bibitem[{Gelfand and Sahu(1999)}]{GelfandSahu99}
Gelfand, A.~E. and Sahu, S.~K. (1999), \enquote{{Identifiability, improper
  priors, and Gibbs sampling for generalized linear models},} \textit{Journal
  of the American Statistical Association}, 94, 247--253.

\bibitem[{Gelfand and Smith(1990)}]{GelfandSmith90}
Gelfand, A.~E. and Smith, A. F.~M. (1990), \enquote{{Sampling-based approaches
  to calculating marginal densities},} \textit{Journal of the American
  Statistical Association}, 85, 398--409.

\bibitem[{Gelman(2006)}]{Gelman06}
Gelman, A. (2006), \enquote{{Prior distributions for variance parameters in
  hierarchical models},} \textit{Bayesian Analysis}, 1, 515--533.

\bibitem[{Geman and Geman(1984)}]{GemanGeman84}
Geman, S. and Geman, D. (1984), \enquote{{Stochastic relaxation, Gibbs
  distributions and the Bayesian restoration of images},} \textit{IEEE
  Transactions on Pattern Analysis and Machine Intelligence}, 6, 721--741.

\bibitem[{Gerber and Furrer(2015)}]{GerberFurrer15}
Gerber, F. and Furrer, R. (2015), \enquote{{Pitfalls in the implementation of
  Bayesian hierarchical modeling of areal count data: An illustration using BYM
  and Leroux models},} \textit{Journal of Statistical Software}, 63, 1--32.

\bibitem[{Golub and Van~Loan(1996)}]{GolubVanLoan96}
Golub, G.~H. and Van~Loan, C.~F. (1996), \textit{{Matrix Computations}},
  Baltimore: The Johns Hopkins University Press, 3rd ed.

\bibitem[{Gonzalez et~al.(2011)Gonzalez, Low, Gretton, and
  Guestrin}]{GonzalezEtAl11}
Gonzalez, J.~E., Low, Y., Gretton, A., and Guestrin, C. (2011),
  \enquote{{Parallel Gibbs sampling: From colored fields to thin junction
  trees},} in \textit{Journal of Machine Learning Research: Proceedings of the
  14th International Conference on Artificial Intelligence and Statistics
  (AISTATS)}, pp. 324--332.

\bibitem[{Grimmett and McDiarmid(1975)}]{Grimmett75}
Grimmett, G.~R. and McDiarmid, C. J.~H. (1975), \enquote{{On colouring random
  graphs},} \textit{Mathematical Proceedings of hte Cambridge Philosophical
  Society}, 33, 313--324.

\bibitem[{Hastings(1970)}]{Hastings70}
Hastings, W. (1970), \enquote{{Monte Carlo sampling methods using Markov chains
  and their application},} \textit{Biometrika}, 57, 97--109.

\bibitem[{Higdon(1998)}]{Higdon98}
Higdon, D.~M. (1998), \enquote{{Auxiliary variable methods for Markov chain
  Monte Carlo with applications},} \textit{Journal of the American Statistical
  Association}, 93, 585--595.

\bibitem[{Hodges et~al.(2003)Hodges, Carlin, and Fan}]{HodgesEtAl03}
Hodges, J.~S., Carlin, B.~P., and Fan, Q. (2003), \enquote{{On the precision of
  the conditionally autoregressive prior in spatial models},}
  \textit{Biometrics}, 59, 317--322.

\bibitem[{Hunziker(2017)}]{HunMapColor17}
Hunziker, P. (2017), \textit{MapColoring: {O}ptimal Contrast Map Coloring}, {R}
  package version 1.0.

\bibitem[{Johnson et~al.(2013)Johnson, Saunderson, and Willsky}]{JohnsonEtAl13}
Johnson, M., Saunderson, J., and Willsky, A. (2013), \enquote{{Analyzing
  Hogwild parallel Gaussian Gibbs sampling},} in \textit{Advances in Neural
  Information Processing Systems 26}, eds. Burges, C. J.~C., Bottou, L.,
  Welling, M., Ghahramani, Z., and Weinberger, K.~Q., Curran Associates, Inc.,
  pp. 2715--2723.

\bibitem[{Kass et~al.(1998)Kass, Carlin, Gelman, and Neal}]{KassEtAl98}
Kass, R.~E., Carlin, B.~P., Gelman, A., and Neal, R. (1998), \enquote{{Markov
  chain Monte Carlo in practice: A roundtable discussion},} \textit{The
  American Statistician}, 52, 93--100.

\bibitem[{Knorr-Held and Rue(2002)}]{KnorrHeldRue02}
Knorr-Held, L. and Rue, H. (2002), \enquote{{On block updating in Markov random
  field models for disease mapping},} \textit{Scandinavian Journal of
  Statistics}, 29, 597--614.

\bibitem[{Koenker and Ng(2016)}]{KoenkerNgSparseM}
Koenker, R. and Ng, P. (2016), \textit{{SparseM: Sparse Linear Algebra}}, {R
  package version 1.74}.

\bibitem[{Krager et~al.(1998)Krager, Motwani, and Sudan}]{KragerEtAl98}
Krager, D., Motwani, R., and Sudan, M. (1998), \enquote{{Approximate graph
  coloring by semidefinite programming},} \textit{Journal of the ACM}, 45,
  246--265.

\bibitem[{Lazar(2008)}]{Lazar08}
Lazar, N.~A. (2008), \textit{{The Statistical Analysis of Functional MRI
  Data}}, New York: {Springer Science + Business Media, LLC}.

\bibitem[{Lee(2013)}]{Lee2013}
Lee, D. (2013), \enquote{{CARBayes: An R package for Bayesian spatial modeling
  with conditional autoregressive priors},} \textit{Journal of Statistical
  Software}, 55, 1--24.

\bibitem[{Lindgren and Rue(2015)}]{LindgrenRue15}
Lindgren, F. and Rue, H. (2015), \enquote{{Bayesian spatial modelling with
  R-INLA},} \textit{Journal of Statistical Software}, 63, 1--25.

\bibitem[{Lindgren et~al.(2011)Lindgren, Rue, and
  Lindstr\"{o}m}]{LindgrenEtAl11}
Lindgren, F., Rue, H., and Lindstr\"{o}m, J. (2011), \enquote{{An explicit link
  between Gaussian fields and Gaussian Markov random fields: The stochastic
  partial differential equation approach},} \textit{Journal of the Royal
  Statistical Society, Series B}, 73, 423--498.

\bibitem[{Liu et~al.(1994)Liu, Wong, and Kong}]{LiuEtAl94}
Liu, J.~S., Wong, W.~H., and Kong, A. (1994), \enquote{{Covariance structure of
  the Gibbs sampler with applications to the comparisons of estimators and
  augmentation schemes},} \textit{Biometrika}, 81, 27--40.

\bibitem[{Liu et~al.(2015)Liu, Kosut, and Willsky}]{LiuEtAl15}
Liu, Y., Kosut, O., and Willsky, A.~S. (2015), \enquote{{Sampling from Gaussian
  Markov random fields using stationary and non-stationary subgraph
  pertubations},} \textit{IEEE Transactions on Signal Processing}, 63,
  576--589.

\bibitem[{Lunn et~al.(2000)Lunn, Thomas, Best, and Spiegelhalter}]{LunnEtAl00}
Lunn, D.~J., Thomas, A., Best, N., and Spiegelhalter, D. (2000),
  \enquote{{WinBUGS -- A Bayesian modelling framework: Concepts, structure, and
  extensibility},} \textit{Statistics and Computing}, 10, 325--337.

\bibitem[{Metropolis et~al.(1953)Metropolis, Rosenbluth, Rosenbluth, Teller,
  and Teller}]{MetropolisEtAl53}
Metropolis, N., Rosenbluth, A.~W., Rosenbluth, M.~N., Teller, A.~H., and
  Teller, E. (1953), \enquote{{Equation of state calculations by fast computing
  machines},} \textit{Journal of Chemical Physics}, 21, 1087--1091.

\bibitem[{Ng and Peyton(1993)}]{NgPetyon93}
Ng, E.~g. and Peyton, B.~W. (1993), \enquote{{Block sparse Cholesky algorithms
  on advanced uniprocessor computers},} \textit{SIAM Journal on Scientific
  Computing}, 14, 1034--1056.

\bibitem[{Niu et~al.(2011)Niu, Recht, R\'{e}, and Wright}]{NiuEtAl11}
Niu, F., Recht, B., R\'{e}, C., and Wright, S.~J. (2011), \enquote{Hogwild! A
  lock-free approach to parallelizing stochastic gradient descent,} in
  \textit{Advances in Neural Information Processing Systems 24}, eds.
  Shawe-Taylor, J., Zemel, R.~S., Bartlett, P.~L., Pereira, F., and Weinberger,
  K.~Q., Curran Associates, Inc., pp. 693--701.

\bibitem[{Polson et~al.(2013)Polson, Scott, and Windle}]{Polson13}
Polson, N.~G., Scott, J.~G., and Windle, J. (2013), \enquote{Bayesian Inference
  for Logistic Models Using P\'{o}lya Gamma Latent Variables,} \textit{Journal
  of the American Statistical Association}, 108, 1339--1349.

\bibitem[{{R Core Team}(2018)}]{R16}
{R Core Team} (2018), \textit{R: A Language and Environment for Statistical
  Computing}, R Foundation for Statistical Computing, Vienna, Austria.

\bibitem[{Robert and Casella(2004)}]{RobertCasella04}
Robert, C. and Casella, G. (2004), \textit{{Monte Carlo Statistical Methods}},
  New York: Springer, 2nd ed.

\bibitem[{Rue(2001)}]{Rue01}
Rue, H. (2001), \enquote{{Fast sampling of Gaussian Markov random fields},}
  \textit{Journal of the Royal Statistical Society, Series B}, 63, 325--338.

\bibitem[{Rue and Held(2005)}]{RueHeld05}
Rue, H. and Held, L. (2005), \textit{{Gaussian Markov Random Fields}}, Boca
  Raton: Chapman \& Hall/CRC.

\bibitem[{Rue et~al.(2009)Rue, Martino, and Chopin}]{RueEtAl09}
Rue, H., Martino, S., and Chopin, N. (2009), \enquote{{Approximate Bayesian
  inference for latent Gaussian models by using integrated nested Laplace
  approximations},} \textit{Journal of the Royal Statistical Society, Series
  B}, 71, 319--392.

\bibitem[{Rue and Tjemland(2002)}]{RueTjem02}
Rue, H. and Tjemland, H. (2002), \enquote{{Fitting Gaussian Markov random
  fields to Gaussian fields},} \textit{Scandinavian Journal of Statistics}, 29,
  31--49.

\bibitem[{Schabenberger and Gotway(2005)}]{SchabenGotway05}
Schabenberger, O. and Gotway, C.~A. (2005), \textit{{Statistical Methods for
  Spatial Data Analysis}}, Boca Raton: Chapman \& Hall/CRC.

\bibitem[{Self et~al.(2018)Self, McMahan, Brown, Lund, Gettings, and
  Yabsley}]{SelfEtAl18}
Self, S. C.~W., McMahan, C.~S., Brown, D.~A., Lund, R.~B., Gettings, J.~R., and
  Yabsley, M.~J. (2018), \enquote{{A large-scale spatio-temporal binomial
  regression model for estimating seroprevalence trends},}
  \textit{Environmetrics}, e2538.

\bibitem[{Sherman(1975)}]{Sherman75}
Sherman, A.~H. (1975), \enquote{{On the efficient solution of sparse systems of
  linear and nonlinear equations},} Unpublished doctoral dissertation, Yale
  University.

\bibitem[{Song et~al.(2008)Song, Fuentes, and Ghosh}]{SongEtAl08}
Song, H.-R., Fuentes, M., and Ghosh, S. (2008), \enquote{{A comparative study
  of Gaussian geostatistical models and Gaussian Markov random fields},}
  \textit{Journal of Multivariate Analysis}, 99, 1681--1697.

\bibitem[{Waller et~al.(1997)Waller, Carlin, Xia, and Gelfand}]{WallerEtAl97}
Waller, L.~A., Carlin, B.~P., Xia, H., and Gelfand, A.~E. (1997),
  \enquote{{Hierarchical spatio-temporal mapping of disease rates},}
  \textit{Journal of the American Statistical Association}, 92, 607--617.

\bibitem[{Xiao et~al.(2009)Xiao, Reilly, and Khodursky}]{XiaoEtAl09}
Xiao, G., Reilly, C., and Khodursky, A.~B. (2009), \enquote{{Improved detection
  of differentially expressed genes through incorporation of gene location},}
  \textit{Biometrics}, 65, 805--814.

\end{thebibliography}

\end{document}